\documentclass{emulateapj}
\usepackage{lscape}

\def\kms{\,km\,s$^{-1}$}
\def\l{\ifmmode\lambda\else$\lambda$\fi}
\def\catwo{Ca{\,\sc ii}}
\def\ctwo{C{\,\sc ii}}
\def\stwo{S{\,\sc ii}}
\def\fetwo{Fe{\,\sc ii}}
\def\cotwo{Co{\,\sc ii}}
\def\sitwo{Si{\,\sc ii}}

\def\naone{Na{\,\sc i}}

\def\dmm{\ifmmode\Delta m_{15}(B)\else$\Delta m_{15}(B)$\fi}
\def\vabs{\ifmmode v_{\rm abs}\else$v_{\rm abs}$\fi}
\def\vpeak{\ifmmode v_{\rm peak}\else$v_{\rm peak}$\fi}

\shorttitle{Line profiles in local and high-$z$ SN~Ia}  
\shortauthors{Blondin et al.}

\begin{document}

\received{2005} 

%%%%%%%%%%%%%%%%%%%%%%%%%%%%%%%%%%%%%%%%%%%%%%%%%%%%%%%%%%%%%%%%%%%
%%
%%   TITLE + FOOTNOTES
%%
%%%%%%%%%%%%%%%%%%%%%%%%%%%%%%%%%%%%%%%%%%%%%%%%%%%%%%%%%%%%%%%%%%%

\title{Using line profiles to test the fraternity of Type
Ia supernovae at high and low redshifts\footnotemark[1]}

\footnotetext[1]{\vspace{0.00cm}Based
in part on observations obtained at the Cerro Tololo Inter-American
Observatory (CTIO), which is operated by the Association of Universities
for Research in Astronomy, Inc. (AURA) under cooperative agreement
with the National Science Foundation (NSF); the European Southern
Observatory, Chile (ESO Programme 170.A-0519); the Gemini Observatory,
which is operated by AURA
under a cooperative agreement with the NSF on behalf
of the Gemini partnership: the NSF (United States), the Particle
Physics and Astronomy Research Council (United Kingdom), the National
Research Council (Canada), CONICYT (Chile), the Australian Research
Council (Australia), CNPq (Brazil), and CONICET (Argentina) (Programs
GN-2002B-Q-14, GN-2003B-Q-11, GS-2003B-Q-11); the Magellan Telescopes
at Las Campanas Observatory; the MMT Observatory, a joint facility of
the Smithsonian Institution and the University of Arizona; and the
F.~L. Whipple Observatory, which is operated by the Smithsonian
Astrophysical Observatory.  Some of the data presented herein were
obtained at the W. M. Keck Observatory, which is operated as a
scientific partnership among the California Institute of Technology,
the University of California, and the National Aeronautics and Space
Administration. The Observatory was made possible by the generous
financial support of the W. M. Keck Foundation.}

\author{
St\'ephane Blondin,\altaffilmark{2} 
Luc Dessart,\altaffilmark{3}
Bruno Leibundgut,\altaffilmark{2}
David Branch,\altaffilmark{4} 
Peter H\"oflich,\altaffilmark{5}
John L. Tonry,\altaffilmark{6}
Thomas Matheson,\altaffilmark{7}
Ryan J. Foley,\altaffilmark{8}
Ryan Chornock,\altaffilmark{8}
Alexei V. Filippenko,\altaffilmark{8}
Jesper Sollerman,\altaffilmark{9} \\
Jason Spyromilio,\altaffilmark{2} 
Robert P. Kirshner,\altaffilmark{10}
W. Michael Wood-Vasey\altaffilmark{10}
Alejandro Clocchiatti,\altaffilmark{11} \\ 
Claudio Aguilera,\altaffilmark{12}
Brian Barris,\altaffilmark{6}
Andrew C. Becker,\altaffilmark{13}
Peter Challis,\altaffilmark{10} 
Ricardo Covarrubias,\altaffilmark{13} \\
Tamara M. Davis,\altaffilmark{14}  
Peter Garnavich,\altaffilmark{15}
Malcolm Hicken,\altaffilmark{10,16}
Saurabh Jha,\altaffilmark{8}
Kevin Krisciunas,\altaffilmark{15} \\
Weidong Li,\altaffilmark{8}
Anthony Miceli,\altaffilmark{13}
Gajus Miknaitis,\altaffilmark{17}
Giuliano Pignata,\altaffilmark{11}
Jose Luis Prieto,\altaffilmark{18}
Armin Rest,\altaffilmark{12} \\
Adam G. Riess,\altaffilmark{19}
Maria Elena Salvo,\altaffilmark{14}
Brian P. Schmidt,\altaffilmark{14}
R. Chris Smith,\altaffilmark{11}
Christopher W. Stubbs,\altaffilmark{10,16} \\
and Nicholas B. Suntzeff\altaffilmark{11} \\
}

\nopagebreak 

\altaffiltext{2}{\vspace{0.00cm}European Southern Observatory,
  Karl-Schwarzschild-Strasse 2, Garching, D-85748, Germany;
  {sblondin@eso.org}, {bleibund@eso.org}, {jspyromi@eso.org}}

\altaffiltext{3}{\vspace{0.00cm}Steward Observatory, 
  University of Arizona, 933 North Cherry Avenue, Tucson, AZ 85721; 
  {luc@as.arizona.edu}}

\altaffiltext{4}{\vspace{0.00cm}Department of Physics and Astronomy, 
  University of Oklahoma, Norman, Oklahoma; {branch@nhn.ou.edu}}

\altaffiltext{5}{\vspace{0.00cm}Department of Astronomy, 
  University of Texas, Austin, TX 78681; {pah@hej1.as.utexas.edu}}

\altaffiltext{6}{\vspace{0.00cm}Institute for Astronomy, University
  of Hawaii, 2680 Woodlawn Drive, Honolulu, HI 96822;
  {barris@ifa.hawaii.edu}, {jt@ifa.hawaii.edu}}

\altaffiltext{7}{\vspace{0.00cm}National Optical Astronomy
  Observatory, 950 N.  Cherry Avenue, Tucson, AZ 85719-4933;
  {matheson@noao.edu}}

\altaffiltext{8}{\vspace{0.00cm} Department of Astronomy,
  University of California, Berkeley, CA 94720-3411;
  {rfoley@astro.berkeley.edu}, {chornock@astro.berkeley.edu},
  {alex@astro.berkeley.edu}, {sjha@astro.berkeley.edu},
  {weidong@astro.berkeley.edu}}

\altaffiltext{9}{\vspace{0.00cm}Stockholm Observatory, AlbaNova,
  SE-106 91 Stockholm, Sweden; {jesper@astro.su.se}}

\altaffiltext{10}{\vspace{0.00cm}Harvard-Smithsonian Center for
  Astrophysics, 60 Garden Street, Cambridge, MA 02138;
  {kirshner@cfa.harvard.edu}, {pchallis@cfa.harvard.edu},
  {mhicken@cfa.harvard.edu}, {cstubbs@fas.harvard.edu},
  {wmwood-vasey@cfa.harvard.edu}} 

\altaffiltext{11}{\vspace{0.00cm}Pontificia Universidad Cat\'{o}lica
  de Chile, Departamento de Astronom\'{i}a y Astrof\'{i}sica, Casilla
  306, Santiago 22, Chile; {aclocchi@astro.puc.cl}}

\altaffiltext{12}{\vspace{0.00cm}Cerro Tololo Inter-American
  Observatory, Casilla 603, La Serena, Chile;
  {caguilera@ctio.noao.edu}, {arest@noao.edu}, {csmith@noao.edu},
  {nsuntzeff@noao.edu}}

\altaffiltext{13}{\vspace{0.00cm}Department of Astronomy, 
  University of Washington, Box 351580, Seattle, WA 98195-1580;
  {becker@darkstar.astro.washington.edu},
  {ricardo@astro.washington.edu}, {amiceli@astro.washington.edu}}

\altaffiltext{14}{\vspace{0.00cm}The Research School of Astronomy and
  Astrophysics, The Australian National University, Mount Stromlo and
  Siding Spring Observatories, via Cotter Rd, Weston Creek PO 2611,
  Australia; {tamarad@mso.anu.edu.au}, {salvo@mso.anu.edu.au},
  {brian@mso.anu.edu.au}} 

\altaffiltext{15}{\vspace{0.00cm}Department of Physics, 
    University of Notre Dame, 
    225 Nieuwland Science Hall, Notre Dame, IN 46556-5670;
  {pgarnavi@nd.edu}, {kkrisciu@nd.edu}}

\altaffiltext{16}{\vspace{0.00cm}Department of Physics, 17 Oxford
  Street, Harvard University, Cambridge MA 02138}

\altaffiltext{17}{\vspace{0.00cm}Department of Physics, 
  University of Washington, Box 351560, Seattle, WA 98195-1560;
  {gm@u.washington.edu}}

\altaffiltext{18}{\vspace{0.00cm}Department of Astronomy, 
  Ohio State University, 4055 McPherson Laboratory, 
  140 W. 18th Ave., Columbus,
  Ohio 43210; {prieto@astronomy.ohio-state.edu}}

\altaffiltext{19}{\vspace{0.00cm}Space Telescope Science Institute,
  3700 San Martin Drive, Baltimore, MD 21218; {ariess@stsci.edu}}

%%%%%%%%%%%%%%%%%%%%%%%%%%%%%%%%%%%%%%%%%%%%%%%%%%%%%%%%%%%%%%%%%%%
%%
%%   ABSTRACT
%%
%%%%%%%%%%%%%%%%%%%%%%%%%%%%%%%%%%%%%%%%%%%%%%%%%%%%%%%%%%%%%%%%%%%

\begin{abstract}
Using archival data of low-redshift ($z<0.01$; CfA and SUSPECT
databases) Type Ia supernovae (SN~Ia) and recent observations of 
high-redshift ($0.16<z<0.64$; \citealt{Matheson/etal:2005}) SN~Ia, 
we study the ``uniformity'' of the spectroscopic properties of nearby
and distant SN~Ia. We find no difference in the measures we
describe here. In this paper, we base our analysis solely on
line-profile morphology, focusing on measurements of the velocity
location of maximum absorption (\vabs) and peak emission (\vpeak). 
Our measurement technique makes it easier to compare low and
high signal-to-noise ratio observations. We also quantify the associated
sources of error, assessing the effect of line blending with
assistance from the parametrized code SYNOW
\citep{Fisher/etal:1999}. We find that the evolution of \vabs\ and
\vpeak\ for  our sample lines (\catwo\ \l3945, \sitwo\ \l6355, and \stwo\
\l\l5454, 5640) is similar for both the low- and high-redshift
samples. We find that \vabs\ for the weak \stwo\ \l\l5454, 5640 lines, and
\vpeak\ for \stwo\ \l5454, can be used to identify fast-declining [$\dmm
> 1.7$] SN~Ia,
which are also subluminous. In addition, we
give the first direct evidence in two high-$z$ SN~Ia spectra of a
double-absorption feature in \catwo\ \l3945, an event also observed,
though infrequently, in low-redshift SN~Ia spectra  
(6/22 SN~Ia in our local sample). Moreover, echoing the recent studies of 
\citet{Dessart/Hillier:2005a,Dessart/Hillier:2005b} in the context of
Type II supernovae (SN~II), we see similar P-Cygni line profiles
in our large sample of SN~Ia spectra. First, the magnitude of
the velocity location at maximum profile absorption may underestimate
that at the continuum photosphere, as observed for example in the
optically thinner line \stwo\ \l5640. Second, we report for the first
time the unambiguous and systematic intrinsic blueshift of peak
emission of optical P-Cygni line profiles in Type Ia spectra, by as
much as 8000\kms. All the high-$z$ SN~Ia analyzed in this paper were
discovered and followed up by the ESSENCE collaboration, and are
now publicly available.
\end{abstract}


%%%%%%%%%%%%%%%%%%%%%%%%%%%%%%%%%%%%%%%%%%%%%%%%%%%%%%%%%%%%%%%%%%%
%%
%%   SECTION 1. -- INTRODUCTION
%%
%%%%%%%%%%%%%%%%%%%%%%%%%%%%%%%%%%%%%%%%%%%%%%%%%%%%%%%%%%%%%%%%%%%

\section{Introduction\label{Sect:intro}}

Type Ia supernovae (SN~Ia) have been the subject of intense
theoretical modeling and dedicated observational campaigns in recent
years. The implications of relative luminosity distance measurements
for low-redshift and high-redshift SN~Ia, namely the requirement for
an additional negative pressure term in Einstein's field equations
(``dark energy''), are extraordinary for cosmologists and particle
physicists alike (\citealt{Riess/etal:1998,Perlmutter/etal:1999}; see
\citealt{Filippenko:2004,Filippenko:2005} for recent reviews). 
These results have not only been confirmed at moderate redshifts
\citep{Tonry/etal:2003,Knop/etal:2003,Barris/etal:2004}, but also at
higher ($z > 1$) redshifts where the universal expansion is in a
decelerating phase \citep{Riess/etal:2004}. Currently, two ongoing
projects are aiming to measure the equation-of-state parameter of the
dark energy: the ESSENCE \citep[Equation of State: SupErNovae 
trace Cosmic Expansion; ][]{Miknaitis/etal:2005, Krisciunas/etal:2005,
Matheson/etal:2005} and SNLS \citep[Supernova Legacy Survey;
][]{Pritchet:2004} projects.

A substantial motivation for
precisely determining potential non-uniformity among high and low-redshift
SN~Ia is how those differences would impact cosmological models. For
example, the use of SN~Ia as distance indicators requires an empirical
relation, verified for nearby SN~Ia, between light-curve width and
maximum luminosity \citep{Phillips:1993}. The extrapolation
of such a relation to SN~Ia at higher redshifts might be inaccurate,
arising possibly from distinct progenitor properties or yet unknown
differences in the explosion mechanism (see
\citealt{Hillebrandt/Niemeyer:2000,Leibundgut:2001,Hoeflich/etal:2003}
for reviews). 

Such evolutionary effects, however, would also need to explain the
apparent brightening of SN~Ia at $z \ga 1$ \citep{Riess/etal:2004}.  
Recently, \citet{Krisciunas/Phillips/Suntzeff:2004} have noted
the absence of a relation between luminosity and light-curve shape in
the near infrared ($JHK$ bands), opening up exciting prospects for
future high-$z$ SN~Ia observations in this (rest-frame) passband. In
the meantime, it is worthwhile to look for potential differences
between SN~Ia at different redshifts, based on their light-curve
properties.

Spectroscopy is better suited than photometry to make
quantitative comparisons between local and high-$z$ SN~Ia. Large
amounts of information are conveyed by spectra on the properties of
the ejecta (chemical composition, velocity/density gradients,
excitation level); subtle differences, blurred together in
photometric measurements, will show up in the spectra.
So far, comparisons of
SN~Ia at different redshifts have only been qualitative in nature
\citep{Coil/etal:2000,Leibundgut/Sollerman:2001}, although preliminary
results on a quantitative 
analysis have been presented by \citet{Lidman:2004}. The ESSENCE
spectra published by \citet{Matheson/etal:2005} clearly show that a
significant fraction of the high-$z$ data is of sufficient quality for
such comparisons, made possible through the public
availability of many local SN~Ia data {\it via} the SUSPECT
database\footnote{SUSPECT: SUpernova SPECTrum Archive,
http://nhn.ou.edu/$\sim$suspect/}. The high-$z$ data, presented by
\citet{Matheson/etal:2005}, are now publicly 
available\footnote{http://www.noao.edu/noao/staff/matheson/spectra.html;
the VLT spectra are also publicly available {\it via} the ESO archive:
http://archive.eso.org/archive/public\_datasets.html}. 

The optical spectra of SN~Ia near maximum light are dominated by
resonance lines of singly ionized, intermediate-mass elements,
Doppler-broadened due to the large expansion velocities in SN~Ia
envelopes; see \citet{Filippenko:1997} for an observational review.
Optically thick lines forming in such fast-expanding ejecta
have a P-Cygni profile shape, characterized essentially by absorption 
blueward of line center and peak emission at line center
\citep{Kirshner/etal:1973}. 
Although these two components always appear qualitatively similar in
P-Cygni profiles associated with optically thick outflows, there
are significant differences, which may become relevant if one seeks
an accurate association with, say, the velocity at the
photosphere -- defined here as the outflow location where the inward
integrated {\it continuum} optical depth is 2/3 -- 
of the SN ejecta \citep{Kirshner/Kwan:1974} or, in another context,
with the asymptotic velocity of a radiatively driven hot star 
wind \citep{Prinja/Barlow/Howarth:1990}.

Recently, \citet{Dessart/Hillier:2005a,Dessart/Hillier:2005b}
performed detailed analyses  
of line-profile formation in SN~II spectra, using hydrogen Balmer 
lines and Fe{~\sc ii} \l5169 diagnostics.
The origin of the possible underestimate of the photospheric
velocity with the use of \vabs\ for optically thinner lines was 
identified by \citet{Branch:1977}, but explained for the first time in
\citet{Dessart/Hillier:2005b}. \citet{Dessart/Hillier:2005a} also
explained the origin of the significant blueshift of P-Cygni profile
emission compared to the rest wavelength of the corresponding line:
for SN~1987A on the 24th of February 1987, this blueshift is on the
order of 7000~km~s$^{-1}$, equivalent to a very sizeable 150~\AA\
\citep[see, e.g.,][]{Dessart/Hillier:2005a}.

Both effects result from the fast declining density distribution in
SN~II, with $n \approx 10$ for a power law given by $\rho(r) = \rho_0 
(R_0/r)^n$, $r$ being the radius and $R_0$ ($\rho_0$) some reference
radius (density). Although the density drop-off in SN~Ia is
estimated to be smaller (see, e.g., \citealt{Hoeflich:1995}), with $n
\approx 7$, this is still large enough for the above two effects to
occur. Such velocity shifts are not trivial: they represent key
observables to constrain the density distribution and the sites of
line formation (and interpret, for example, line-polarization measurements),
the magnitude of disk-occultation and continuum optical-depth effects,
and the ubiquitous but modulated influence of line overlap. The velocity 
locations corresponding to maximum absorption and peak emission are
thus carriers of important information on the SN ejecta; they are,
moreover, convenient and well-defined observables, and can thus be
used to objectively compare SN~Ia at different redshifts. The present
paper is the result of such an investigation, using 229 spectra of
local ($z < 0.05$) and 48 of high-$z$ ($0.16 < z < 0.64$) SN~Ia, at
phases between $-2$ weeks to +3 weeks from maximum light.  

To minimize measurement biases, introduced
primarily by the signal-to-noise ratio (S/N) obtained for the faint,
high-redshift SN~Ia, we develop a spectral-smoothing technique, which
takes account of the
expected large widths of observed SN~Ia spectral features.
We give a detailed account of our measurement technique and
associated error model in Sect.~2. The results of these
measurements are presented in Sect.~3, with 
individual discussion of \vabs\ and \vpeak\ for \catwo\ \l3945, 
\sitwo\ \l6355, and \stwo\ \l\l5454, 5640.
We discuss the wide range of \vabs\ values found for the different lines,
the large magnitude of \vpeak\ for the optically thinner 
lines \stwo\ \l\l5454,5640, and the detection, for the first time, of
double-absorption features in \catwo\ \l3945 in high-$z$ SN~Ia
spectra. We provide insights into the nature of the above measurements
by illustrating, following
\citet{Dessart/Hillier:2005a,Dessart/Hillier:2005b}, some aspects of
line and continuum formation in SN~Ia spectra; namely, we explain the
origin of the blueshift of peak emission and the relation of the
absorption velocity to the photospheric and expansion velocities. In
Sect.~4, we present our conclusions.

%%%%%%%%%%%%%%%%%%%%%%%%%%%%%%%%%%%%%%%%%%%%%%%%%%%%%%%%%%%%%%%%%%%
%%
%%   SECTION 2 -- MEASUREMENT TECHNIQUES
%%
%%%%%%%%%%%%%%%%%%%%%%%%%%%%%%%%%%%%%%%%%%%%%%%%%%%%%%%%%%%%%%%%%%%

\section{Measurement techniques\label{Sect:measurementtechniques}}

Any comparison between low-$z$ and high-$z$ SN~Ia spectra suffers
from the significantly degraded signal quality for the latter,
due to the limited integration time available per spectrum when
undertaking large SN~Ia surveys \citep{Matheson/etal:2005}.
To minimize this bias, we have developed a smoothing technique, presented
in detail in the following section.

\subsection{Smoothing Supernova Spectra\label{Sect:filter}}

We apply a filter to both local and
high-$z$ SN~Ia spectra, to measure the absorption and emission-peak
velocities in a consistent manner. This filter takes into account the
wavelength-dependent nature of the noise in optical  ground-based
spectra, and is based on the fact that supernova spectral features are
intrinsically broadened due to the large expansion velocity in the
corresponding line-formation region. Assuming this broadening to have
typical values of $\sim$1000--3000\kms, one can write down a
``smoothing factor'' $d\lambda/\lambda$ as ($c$ is the speed of light
in vacuum)

\begin{equation}
\Delta v_{\rm line} \approx 0.003-0.01 c \quad \Longrightarrow \quad
\frac{d\lambda}{\lambda} \approx 0.003-0.01. 
\end{equation}
 
\noindent
For such a spectrum, a well-suited filter is a Gaussian of the same
width, $\sigma_{\rm g}$. This assumes the S/N of the spectrum 
to be uniform over the whole wavelength range,
which is clearly not the case for ground-based observations in which
sky emission lines increase the noise in the red part 
of the spectrum. Thus, we instead adopt an inverse-variance weighted
Gaussian filter.

Let $\overrightarrow{F_{\rm SN}}$ and $\overrightarrow{F_{\rm var}}$
be the 1D, flux-calibrated supernova spectrum and its 
corresponding variance spectrum (usually the variance of the
optimally extracted spectrum; see \citealt{Horne:1986}). Both spectra
share the same wavelength axis, $\overrightarrow{\l}$. At each
wavelength element $\lambda_i$, construct a Gaussian
$\overrightarrow{G_i}$ of width 

\begin{equation}
\sigma_{{\rm g},i} = \lambda_i \frac{d\lambda}{\lambda}.
\end{equation}

\noindent
We thus have

\begin{equation}
\overrightarrow{G_{i}} = 
\left( 
\begin{array}{c} 
G_{i,1} \\
G_{i,2} \\
\vdots  \\
G_{i,N_{l}}
\end{array}
\right) = \frac{1}{\sqrt{2\pi}} 
\exp \left[-
\frac{1}{\sigma_{{\rm g},i}}
\left(
\begin{array}{c}
\lambda_1 - \lambda_i \\
\lambda_2 - \lambda_i \\
\vdots                \\
\lambda_{N_{l}} - \lambda_i
\end{array}
\right)
\right]^2
\end{equation}

\noindent
where $N_{l}$ is the number of wavelength elements of a subset of
$\overrightarrow{\l}$, centered on $\l_i$. The inverse-variance
weighted Gaussian is then

\begin{equation}
\overrightarrow{W_{i}} = 
\left( 
\begin{array}{c} 
W_{i,1} \\
W_{i,2} \\
\vdots  \\
W_{i,N_{l}}
\end{array}
\right) = 
\left(
\begin{array}{c}
G_{i,1}/F_{{\rm var},1} \\
G_{i,2}/F_{{\rm var},2} \\
\vdots \\
G_{i,N_{l}}/F_{{\rm var},N_{l}} \\
\end{array}
\right)
\end{equation}

\noindent
and the corresponding smoothed flux at $\lambda_i$ is therefore 

\begin{equation}
F_{{\rm TS},i} = \frac{\displaystyle \sum_{j} W_{i,j} F_{{\rm SN},j} }
{\displaystyle \sum_{j} W_{i,j} }.
\end{equation}

\noindent
By repeating this process for each wavelength element $\lambda_i$, we
obtain the smoothed supernova spectrum $\overrightarrow{F_{\rm TS}}$. 

\begin{figure}	
\epsscale{1.2}
\plotone{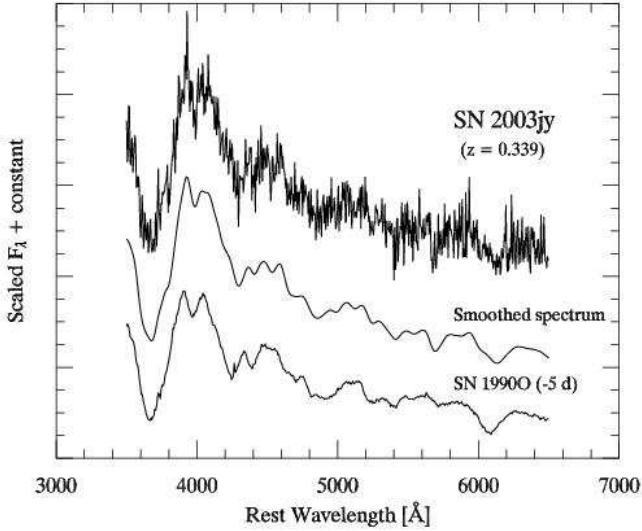}
\caption{
({\it From top to bottom}) A high-$z$ SN~Ia spectrum
(SN~2003jy; $z=0.339$), its smoothed version (with
$d\lambda/\lambda=0.005$), and a local zero-redshift template (SN~1990O at
$-5$~d), plotted for comparison with the smoothed spectrum.
\label{Fig:filter}}
\end{figure}

 We show the result of applying this spectral smoothing algorithm
with $d\lambda/\lambda = 0.005$ in Fig.~\ref{Fig:filter}. The upper
spectrum is that of SN~2003jy ($z=0.339$), plotted 
as a function of rest wavelength. 
The spectrum just below is the smoothed version, showing
how most of the high-frequency noise has been removed, while the
lower-frequency SN spectral features have been preserved. In this
paper, we use the same smoothing factor $d\lambda/\lambda = 0.005$ in
applying this smoothing technique to both the local and high-$z$ SN~Ia
spectra.

%%%%%%%%%%%%%%%%%%%%%%%%%%%%%%%%%%%%%%%%%%%%%%%%%%%%%%%%%%%%%%%%%%%
%
%    Fig. -- filter
%
%%%%%%%%%%%%%%%%%%%%%%%%%%%%%%%%%%%%%%%%%%%%%%%%%%%%%%%%%%%%%%%%%%%

\begin{figure}	
\epsscale{1.2}
\plotone{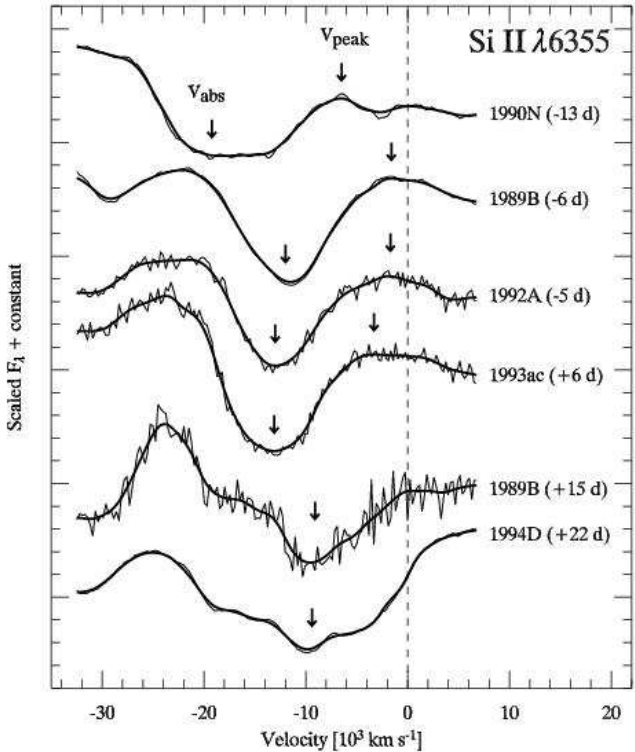}
\caption{
P-Cygni profiles of \sitwo\ \l6355 in local SN~Ia. Overplotted
on the spectra are their smoothed version (thick black line, see
Sect.~\ref{Sect:filter}). The
arrows indicate the measured maximum absorption and emission peak, used to
determine \vabs\ and \vpeak. Contamination of the sides
of the \sitwo\ absorption profile by strengthening emission/absorption 
from iron-group elements is apparent at late phases ($\gtrsim 2$ weeks),
and prevents us from measuring the emission-peak velocity.
\label{Fig:pcygniprofiles}}
\end{figure}

We show examples of smoothed spectra of several local SN~Ia at
different phases (and with different S/N) in
Fig.~\ref{Fig:pcygniprofiles}, where we concentrate on the \sitwo\ \l6355
feature, plotted in velocity space (assuming
$\lambda_0=6355$~\AA\ and using the relativistic Doppler formula). We
also show (down arrows) the location of maximum absorption and
emission peak, measured with our smoothing technique and corresponding
to \vabs\ and \vpeak. We see that line profiles in SN~Ia come in many
shapes and sizes, from the well-defined absorption trough of SN~1992A
at $-5$~d to the flat-bottomed one of SN~1990N at $-13$~d, which
extends over $\ga$ 5000\kms\ (perhaps due to contamination from \ctwo\
\l6580; see \citealt{Fisher/etal:1997,Mazzali:2001}). The
emission-peak region is often less well defined than the absorption
trough \citep{Jeffery/Branch:1990}, and is more affected by
contamination by emission from iron-group elements at late phases
($\ga 2$ weeks past maximum brightness). Our spectral smoothing technique does a
fine job in reproducing the broad features in SN~Ia spectra, both for
low-S/N spectra and for contaminated line profiles.

%%%%%%%%%%%%%%%%%%%%%%%%%%%%%%%%%%%%%%%%%%%%%%%%%%%%%%%%%%%%%%%%%%%
%
%    Fig. -- pcygniprofiles
%
%%%%%%%%%%%%%%%%%%%%%%%%%%%%%%%%%%%%%%%%%%%%%%%%%%%%%%%%%%%%%%%%%%%

For some spectra we do not have a corresponding variance spectrum at
our disposal, either because these variance spectra are not archived in
the spectral databases (this is the case for many local SN~Ia spectra),
or because the pipeline used to reduce the spectra did not
generate the variance spectra (as was the case for spectra taken with
Keck+LRIS, at the time of data reduction). For these spectra we use a
fiducial sky spectrum taken with the same telescope/instrument
combination in place of a variance spectrum. This is adequate since
our smoothing technique relies on the relative variance only,
and we expect sky emission to be the dominant source of noise in
our optical ground-based spectra. 

We then spline interpolate the smoothed spectrum
$\overrightarrow{F_{\rm TS}}$ onto a $0.1$~\AA\ resolution grid, and
determine the wavelengths of maximum absorption ($\lambda_{\rm abs}$ )
and emission peak ($\lambda_{\rm peak}$).
The absorption and emission-peak
velocities are then calculated using the relativistic Doppler formula,

\begin{equation}
v_{\rm Doppler, rel} = c \left\{ \frac{{[(\Delta\lambda/\lambda_0)+1]}^2 -
1}{{[(\Delta\lambda/\lambda_0)+1]}^2 + 1} \right\}, 
\end{equation}

\noindent
where $\Delta\lambda = \lambda_{\rm abs} - \lambda_0$ or
$\Delta\lambda = \lambda_{\rm peak} - \lambda_0$ when inferring \vabs\
or \vpeak, respectively, and $\lambda_0$ is the rest-frame wavelength
of the corresponding transition.

\begin{figure}	
\epsscale{1.2}
\plotone{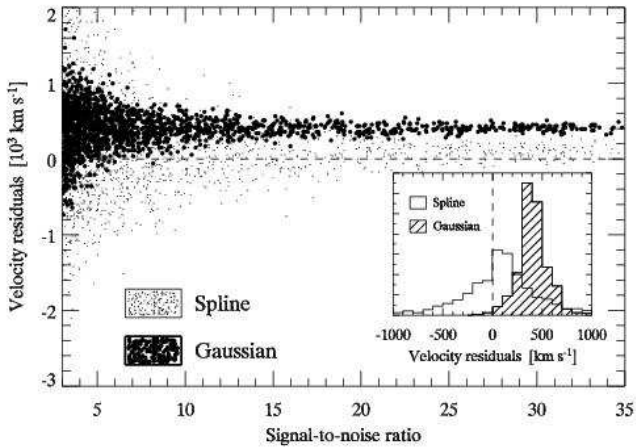}
\caption{Velocity residuals when fitting the minimum of the P-Cygni
profile of \sitwo\ \l6355 in 1989B at $-$6~d with a Gaussian ({\it black
dots}), or when determining the minimum from a spline interpolation of an
inverse-variance weighted Gaussian spectrum ({\it small points}).
The original S/N ratio of the spectrum is $\sim$70. Each of the
points corresponds to the original spectrum to which we have added
random Poisson noise weighted by a fiducial sky spectrum. The inset
shows the distribution of velocity residuals for both the spline ({\it
empty}) and Gaussian ({\it hashed}) techniques.
\label{Fig:min_s2n}}
\end{figure}

This approach has the advantage over using a Gaussian fit to the
overall absorption/emission profiles, since it makes no assumption on
their shape (in particular, whether the absorption/emission profiles are
symmetric or not). Moreover, fitting a function to the whole $\sim100-200$~\AA-wide 
profiles would enhance the impact of line overlap on
the fit. In Fig.~\ref{Fig:min_s2n} we show velocity residuals as a
function of S/N, when using a Gaussian fit to the absorption
profile, and when spline interpolating a spectrum smoothed using our
algorithm. Here the ``noise'' is defined as the root-mean-square (RMS)
deviation between the input spectrum 
$\overrightarrow{F_{\rm SN}}$ and the filtered spectrum
$\overrightarrow{F_{\rm TS}}$, with $d\lambda/\lambda = 0.005$: 

\begin{equation}
{\rm S/N\ } = \frac{ |\overline{\overrightarrow{F_{\rm SN}}}| } 
{\sqrt{(1/N_{\lambda})  \displaystyle \sum_{j=1}^{N_{\lambda}} (
|\overrightarrow{F_{{\rm SN},j}}| - |\overrightarrow{F_{{\rm TS},j}}|
)^2}},  
\end{equation}

\noindent
where $N_{\lambda}$ is the number of elements in
$\overrightarrow{\l}$. 
We find this to be an accurate description of the actual mean
S/N, which then enables us to evaluate the mean S/N of spectra for
which we do not have the corresponding variance spectrum. 

Each of the points of Fig.~\ref{Fig:min_s2n} corresponds
to a fit to the absorption profile of \sitwo\ \l6355 in SN~1989B at
$-6$~d (S/N $\approx 70$ in that spectral region), for which we have added
increasing random Poisson noise weighted by a fiducial sky
spectrum, to reproduce signal-to-noise ratios in the range 2--40. 
In this case, we are making a systematic error of $\sim+400$\kms\ 
when using a Gaussian, while this uncertainty drops to
$\lesssim 100$\kms\ for the spline interpolation method. The latter
method is more sensitive to the S/N, namely the drop in precision with
decreasing S/N is more significant for the spline ($\sigma_{\rm
spline} \approx 320$\kms) than for the Gaussian ($\sigma_{\rm gauss}
\approx 130$\kms). We use such simulations to evaluate the error due to
a spectrum's S/N on our inferred values of \vabs\ and \vpeak\ (see
next Sect.). All the measurements in this paper make use of the
spline-interpolation method.

%%%%%%%%%%%%%%%%%%%%%%%%%%%%%%%%%%%%%%%%%%%%%%%%%%%%%%%%%%%%%%%%%%%
%
%    Fig. -- min_s2n
%
%%%%%%%%%%%%%%%%%%%%%%%%%%%%%%%%%%%%%%%%%%%%%%%%%%%%%%%%%%%%%%%%%%%
%
% Figure showing the velocity residual for the Si II 6355 line 
% in 1989B at -7d when fitting the minimum with a Gaussian and
% a spline, as a function of signal-to-noise ratio
%

\subsection{Error Budget\label{Sect:errorbudget}}

In this section we give a detailed account of the various sources of
systematic error. We give estimates of
these errors in Table \ref{Table:errors}. The elaboration of an error
model is an important part of the comparison of \vabs\ and \vpeak\
amongst local SN~Ia, and between the local and high-$z$ SN
Ia. Several authors presenting measurements of \vabs\ in local SN~Ia 
either do not give their error model \citep{Benetti/etal:2005} or do
not report errors at all \citep{Patat/etal:1996}.

Note that we do not include errors related to line blending, as this
would require detailed modeling of every individual spectrum. We
discuss the issue of line blending when presenting our results in
Sect.~3.

%%%%%%%%%%%%%%%%%%%%%%%%%%%%%%%%%%%%%%%%%%%%%%%%%%%%%%%%%%%%%%%%%%%
%
%    Table -- Error budget
%
%%%%%%%%%%%%%%%%%%%%%%%%%%%%%%%%%%%%%%%%%%%%%%%%%%%%%%%%%%%%%%%%%%%
%
% Table summarizing error budget
%
\begin{deluxetable*}{l l}
\tablewidth{0pt}
\tablecaption{Measurement errors \label{Table:errors}}
\tablehead{
\colhead{Error source}            & 
\colhead{Error size [\kms]} 
}   

\startdata
\multicolumn{2}{l}{\it Errors related to SN~Ia spectral properties:} \\     
Line blending/contamination        & Not included                              \\
Contamination from host galaxy     & Not included   \\
\hline
\multicolumn{2}{l}{\it Errors related to the measurement:}                                                 \\
Skewed profile                     & $\sim1000$   \\
Signal-to-noise ratio              & $\sim[200,500,700,1200]$ for S/N $\approx [20,10,5,2]$ per 5~\AA\ bin \\
\hline
\multicolumn{2}{l}{\it Other errors:}                                                                      \\
Redshift uncertainty               & $\sim[200,3000]$ for galaxy and SN redshift, respectively   \\
Reddening uncertainty $\sigma_{E(B-V)}$ & $\lesssim 100$ for $\sigma_{E(B-V)} \lesssim 0.5$ mag \\
Use of classical Doppler formula   & $\sim[200,400,600]$ for $v\approx [10,15,20]\times10^3$\kms              \\
\enddata 

\end{deluxetable*}

\subsubsection{Errors Related to the Measurement} 

The smoothing and spline-interpolation technique we use to measure the
absorption and emission-peak velocities minimizes the impact of
inherently asymmetric profiles. However, in some cases we are unable
to rely entirely on this method. This occurs when (i) the profile is
highly skewed over a wavelength scale $\Delta\lambda/\lambda <
d\lambda/\lambda$ ($d\lambda/\lambda = 0.005$ here); (ii) the feature
is too weak, yet a minimum is clearly identifiable; (iii) the
absorption profile has a flat minimum (e.g., \sitwo\ \l6355 in SN~1990N
at early phases; Fig.~\ref{Fig:pcygniprofiles}); (iv) there is a sharp
feature which affects the smoothing technique  
(galaxy line, cosmic ray, noise spike). For (i) and (ii) we resort to
a smaller smoothing factor ($0.001 \le d\l / \l \le 0.003$).
For (iii), we report the velocity corresponding
to the blue edge (i.e., the optically thinner part) of the
absorption profile, and associate a lower 
error bar corresponding to the velocity difference between the blue
and red edges of the profile. For (iv), we simply use a linear
interpolation over the sharp feature when its width is less than
$(d\lambda/\lambda) \lambda_{\rm abs}$ (for \vabs\ measurements) or
$(d\lambda/\lambda) \lambda_{\rm peak}$ (for \vpeak\ measurements). 

The S/N of the input spectrum will limit the
accuracy of the measurement. For every local SN~Ia spectrum in our
database, we progressively degrade the S/N of the spectrum by adding
random Poisson noise weighted by a fiducial sky spectrum (see previous
Sect.), and construct plots of velocity residuals {\it vs.} S/N (as in
Fig.~\ref{Fig:min_s2n}) for each of the four spectral features
studied here. Most of the local SN~Ia spectra used in this study have
S/N $> 10$ (per 5~\AA\ bin), and the associated error is $\la
500$\kms\ (Table \ref{Table:errors}), whereas for many of the
high-$z$ spectra, the error due to poor S/N can be $> 1000$\kms.

\subsubsection{Further Errors \label{furthererrors}}

Further errors affecting measurements of \vabs\ and \vpeak\ include
the following.

\begin{itemize}

\item[1.] {{\it Redshift of parent galaxy:} All the supernova
spectra have been corrected for the heliocentric velocity of their
host galaxy, $z_{\rm gal}$. This measurement is affected not only
by the galaxy's internal velocity dispersion ($\sim[200,100]$\kms\
for [early,late]-type galaxies; \citealt{McElroy:1995}), but
also by the position of the SN within the galaxy. An
illustration of this is SN~1994D in NGC~4526 for which the NASA/IPAC
Extragalactic Database (NED) gives $cz=+448$\kms\ but
\citet{King/etal:1995} measure $cz=+880$\kms, {\it at the supernova
position}. We have therefore added a $\sigma_{cz,{\rm gal}} \approx
200$\kms\ error in quadrature to the total error to account 
for this effect. Note that for some of the high-$z$ SN~Ia the
redshift has been determined {\it via} cross-correlation with local SN~Ia
spectral templates, with a typical error $\sigma_{z,{\rm SN}} \approx
0.01 \equiv 3000$\kms\ \citep{Matheson/etal:2005}. Thus, for these
high-$z$ SN the major source of error is due to redshift. Note that
such an error leads to a global shift in velocity, and is different in
nature from effects such as line overlap which are difficult to assess
and vary from line to line.}

\item[2.] {{\it Reddening:} All of the local and high-$z$ SN~Ia
spectra have been corrected for both host galaxy and Galactic
\citep{Schlegel/Finkbeiner/Davis:1998} reddening. For the local
SN~Ia, we use the host-galaxy reddenings of
\citet{Phillips/etal:1999} (see Table \ref{Table:localsn1adata}). For
the high-$z$ SN~Ia, we use the reddening values derived from fitting
the light curves using the algorithm of
\citet{Prieto/Rest/Suntzeff:2005}.
An error in the reddening correction applied to a spectrum can affect the
overall continuum slope and in turn bias \vabs\ to
higher(lower) values, if the reddening is under(over)-estimated. We
have run a simulation to evaluate the errors associated with
reddening mis-estimates. They are $\la 50$\kms\  for
$\sigma_{E(B-V)} \la 0.3$ mag. Since the reddening is typically
known down to $\sim0.1$ mag in local and high-$z$
SN~Ia (see Tables \ref{Table:localsn1adata} \&
\ref{Table:hzsn1adata}), we ignore this error.}

\item[3.] {{\it Host-galaxy contamination:}  
We do not include errors related to contamination of SN~Ia spectra by
host-galaxy light. The S/N of most local SN~Ia
spectra is such that the line shapes are expected to be little
affected by host-galaxy contamination. For the high-$z$ data, we
cannot make this assumption. Nor do we have a reliable way of
evaluating the amount of  galaxy light present in our spectra, and
we ignore this error. All the VLT spectra analyzed
here and presented by \citet{Matheson/etal:2005} were extracted using
a 2D deconvolution technique employing the Richardson-Lucy restoration
method, minimizing the contamination from the host galaxy
\citep{Blondin/etal:2005}. However, we see no systematic effect
between the VLT spectra and those from other telescopes.}

\end{itemize}

We further assume the wavelength
calibration to be accurate to within $\pm 0.5$~\AA, which corresponds
to errors $< 50$\kms\ (not included in the error budget). 

We note that, since absorption velocities in SN~Ia ejecta can
reach $\sim30,000$\kms\ ($0.1c$), relativistic corrections
to the classical Doppler formula become noticeable ($\sim1500$\kms). 
Different authors may or may not apply the relativistic
Doppler formula in their determination of absorption velocities, and
systematic differences can result from a blind comparison of different
measurements. All the velocities calculated here make use of the
relativistic Doppler formula -- this is not strictly consistent 
when comparisons are made with model atmosphere computations that
do not take explicitly into account the relativistic terms in the 
radiative transfer equation.  

An error in the SN phase is of a different nature as it is not a
direct error on a velocity measurement, but rather an indirect one
affecting the correlation of \vabs\ and \vpeak\ measurements with SN
phase. The typical error in the time of
rest-frame $B$-band maximum is $\lesssim 1$ day (Table 
\ref{Table:localsn1adata}). This error will of course
directly propagate as an error in the phase of the supernova
spectrum. For the local SN~Ia, we use an updated version of the
multi-color light-curve shape (MLCS) method of
\citet{Riess/Press/Kirshner:1996}, MLCS2k2
\citep{Jha:2002,Jha/Riess/Kirshner:2005}, to determine the time of
$B$-band maximum. We could use the MLCS2k2 $1\sigma$ error on
HJD$_{\rm max}$ as the $1\sigma$ error on the SN phase, but the
MLCS2k2 templates are sampled only once per day, introducing sampling
errors on the order of $\sim0.5$ day. For the local SN~Ia we add a
fiducial $\pm0.5$ day error in quadrature to the MLCS2k2 error in
HJD$_{\rm max}$, while for the high-$z$ ones we use the $1\sigma$
error output by the light-curve fitting routine of
\citet{Prieto/Rest/Suntzeff:2005}.

%%%%%%%%%%%%%%%%%%%%%%%%%%%%%%%%%%%%%%%%%%%%%%%%%%%%%%%%%%%%%%%%%%%
%%
%%   SECTION 3 -- Results
%%
%%%%%%%%%%%%%%%%%%%%%%%%%%%%%%%%%%%%%%%%%%%%%%%%%%%%%%%%%%%%%%%%%%%

\section{Results\label{Sect:results}}

In this section,  using the above method, we present absorption
(Sect.~\ref{Sect:vabs}) and 
emission-peak (Sect.~\ref{Sect:vpeak}) velocity measurements for the
\catwo\ \l3945, \sitwo\ \l6355, and \stwo\ \l\l5454, 5640 
line profiles (Table \ref{Table:characteristicwavelengths}). 
These lines do not have the same observed profile shape, 
presumably because they form differently, and have the potential 
to reveal distinct aspects of the SN outflow.
Note that here, these measurements are sometimes compared
with the velocity at the photosphere; let us stress again that
throughout this paper, we refer to the photosphere as the outflow
location where the inward integrated {\it continuum} optical depth is
2/3 -- no account is made of line opacity in this definition.

Our sample 
comprises 30 local SN~Ia with phases between $-14$~d and +30~d from
$B$-band maximum (Table \ref{Table:localsn1adata}), and 37 high-$z$
($0.16<z<0.64$) SN~Ia with (rest-frame) phases between $-12$~d and
+19~d from (rest-frame) $B$-band maximum (Table
\ref{Table:hzsn1adata}; \citealt{Matheson/etal:2005}). 
We thus performed measurements for a total of 229 local 
and 48 high-$z$ spectra. Emission-peak velocity measurements, more
affected by line contamination (see \citealt{Jeffery/Branch:1990}), 
are only quoted for 178 local and 39 high-$z$ spectra. 
In Tables~\ref{Table:vabslocal}--\ref{Table:vpeakhighz}, we present
the relativistic Doppler velocities \vabs\ and \vpeak\ corresponding to the
above four line diagnostics, using for each the rest-frame wavelength given in 
Table~\ref{Table:characteristicwavelengths}.
For the doublet lines of \catwo\ and \sitwo\, this corresponds
to the $gf$-weighted mean wavelength, where $g$ and $f$ are the
statistical weight and oscillator strength of the transition,
respectively.  For the two \stwo\ features we instead use the wavelength
of the highest $\log(gf)$ transition, due to the large number of
transitions involved (see Sect.~\ref{Sect:vabsobs}).

%%%%%%%%%%%%%%%%%%%%%%%%%%%%%%%%%%%%%%%%%%%%%%%%%%%%%%%%%%%%%%%%%%%
%
%    Table -- Atomic line data
%
%%%%%%%%%%%%%%%%%%%%%%%%%%%%%%%%%%%%%%%%%%%%%%%%%%%%%%%%%%%%%%%%%%%
%
% Table of atomic data and charateristic wavelengths of transitions
%
\begin{center}
\begin{deluxetable*}{l l c l c c}
\tabletypesize{\scriptsize}
\tablewidth{0pt}
\tablecaption{Characteristic wavelengths of atomic
transitions{\tablenotemark{a}}
\label{Table:characteristicwavelengths}}  
\tablehead{
\colhead{Ion}      & 
\colhead{Multiplet designation} & 
\colhead{\l\ [\AA]}    & 
\colhead{$\log(gf)$}         & 
\colhead{$\lambda_{gf}$ [\AA]} &
\colhead{$\lambda_{\rm used}$ [\AA]{\tablenotemark{b}}}}   
\startdata                  
\catwo                   & 4s$^2$S--4p$^2$P$^0$                       & 3933.66, 3968.47 & 0.134, $-$0.166  & 3945.28 & 3945 \\ 
\stwo{\tablenotemark{c}} & 4s$^4$P--4p$^4$D$^0$                       & 5432.80, 5453.86 & 0.311, \phs0.557 & 5442.69 & 5454 \\ 
\stwo{\tablenotemark{d}} & 3d$^4$F--4p$^4$D$^0$, 4s$^2$P--4p$^2$D$^0$ & 5606.15, 5639.98 & 0.156, \phs0.330 & 5638.12 & 5640 \\
\sitwo                   & 4s$^2$S--4p$^2$P$^0$                       & 6347.11, 6371.37 & 0.297, $-$0.003  & 6355.21 & 6355 \\
\enddata
\tablenotetext{a}{From \citet {Kurucz/Bell:1995}.}   
\tablenotetext{b}{Assumed rest-frame wavelength for the transition of
the given ion; for \catwo\ and \sitwo\ this is simply the $gf$-weighted
mean wavelength of the two strongest transitions, while for \stwo\ this
is the wavelength corresponding to the strongest transition.}
\tablenotetext{c}{There are five transitions corresponding to \stwo\ in
the range $5400$~\AA$ < \lambda < 5500$~\AA; we list the two
corresponding to the largest $gf$ values. Note that the value for
$\lambda_{gf}$ corresponds to the $gf$-weighted mean of these five
transitions.} 
\tablenotetext{d}{There are eight transitions corresponding to \stwo\ in
the range $5600$~\AA$ < \lambda < 5700$~\AA; we list the two
corresponding to the largest $gf$ values. Note that the value for
$\lambda_{gf}$ corresponds to the $gf$-weighted mean of these eight
transitions.}   
\end{deluxetable*}
\end{center}

\placetable{Table:localsn1adata}

\placetable{Table:hzsn1adata}

To facilitate the visual inspection of figures,
we show the temporal evolution of \vabs\ 
(Figs.~\ref{Fig:vabsca2red}, \ref{Fig:vabssi2}, and \ref{Fig:vabss2})
and \vpeak\ 
(Figs.~\ref{Fig:vpeakca2}--\ref{Fig:vpeaks2blue_norm2s2red}) for all
spectra by grouping data points according to the decline-rate
parameter \dmm\ (the decline in $B$-band magnitudes between maximum 
-- +0~d -- and +15~d; \citealt{Phillips:1993,Phillips/etal:1999}) 
of the corresponding SN~Ia. Following this selection criterion, our sample of
local (high-$z$) SN~Ia  has 10 (17) objects with $\dmm < 1.0$, 17 (20)
with $1.0 \le \dmm \le 1.7$, 3 (0) with $\dmm > 1.7$; the lack
of $\dmm > 1.7$ SN~Ia in our high-$z$ sample could be due to a
selection effect \citep{Miknaitis/etal:2005,Krisciunas/etal:2005}.
Note that when computing the decline rate \dmm\ of high-$z$ SN~Ia, 
time dilation is accounted for by scaling the time axis by a factor
$(1+z)^{-1}$ \citep{Leibundgut/etal:1996,Goldhaber/etal:2001}.

%%%%%%%%%%%%%%%%%%%%%%%%%%%%%%%%%%%%%%%%%%%%%%%%%%%%%%%%%%%%%%%%%%%
%%
%%   Absorption velocities
%%
%%%%%%%%%%%%%%%%%%%%%%%%%%%%%%%%%%%%%%%%%%%%%%%%%%%%%%%%%%%%%%%%%%%

\subsection{Absorption Velocities\label{Sect:vabs}}

All {\it spectroscopic} velocity measurements reported in this paper
are negative, and thus correspond to blueshifts; to avoid any confusion,
we apply the standard rules of arithmetic and say, e.g., that a 
\vabs\ measurement {\it increases} from $-$25,000\kms\ to $-$15,000\kms, 
while the simplistic interpretation of such a variation suggests that 
the corresponding location of maximum absorption {\it decreases} from 
outflow {\it kinematic} velocities of 25,000\kms\ down to 
15,000\kms\ -- see Sect.~\ref{Sect:vabscmf}.

\subsubsection{Presentation of \vabs\ Measurements\label{Sect:vabsobs}}

\begin{figure}	
\epsscale{1.1}
\plotone{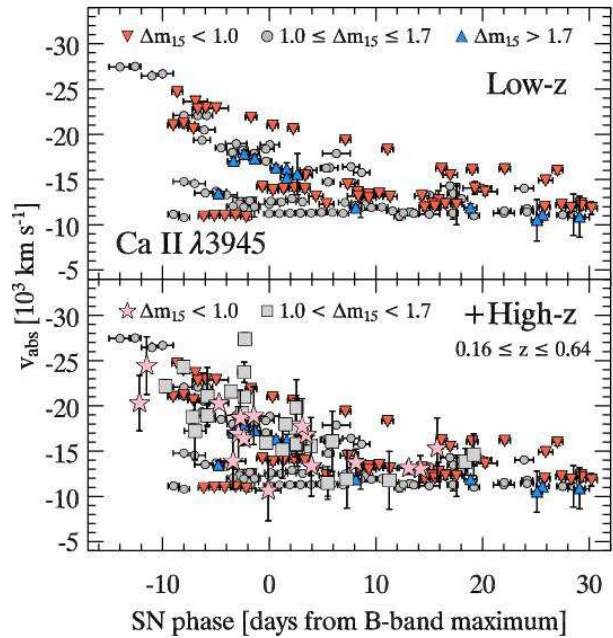}
\caption{{\it Upper panel:} Absorption velocities for \catwo\ \l3945 
in local SN~Ia, for three different \dmm\ ranges. If a double-absorption 
is present, only the redder component is plotted. 
{\it Lower panel:} The high-$z$ data are overplotted.
See the electronic edition of \aj\ for a color version of this figure.
\label{Fig:vabsca2red}}
\end{figure}

In the top panel of Fig.~\ref{Fig:vabsca2red}, we show \vabs\
measurements for the local SN~Ia sample for the \catwo\ \l3945 feature,
as a function of phase and ordered according to their decline rate:
$\dmm<1.0$ (``slow-decliners,'' {\it downward-pointing triangles}),
$1.0<\dmm<1.7$ ({\it circles}), and $\dmm>1.7$ (``fast-decliners,''
{\it upward-pointing triangles}). For
the few objects showing a double-absorption \catwo\ feature (tagged
``blue'' and ``red'' according to wavelength; see Sect.~\ref{Sect:hvca2} and
Table \ref{Table:vabslocal}), we plot only the redder component. 
We invert the ordinate (\vabs) axis for consistency with previously 
published measurements, which usually associate the absorption blueshifts 
with positive velocities.

The absorption velocities for \catwo\ \l3945 reveal two \vabs\
sequences at pre-maximum phases: one sequence shows a steady increase 
in \vabs\ from a minimum of $\ga -25,000$\kms\ at the earliest
observed phases ($\la -10$~d) to $\sim-15,000$\kms\ at $B$-band
maximum, after which the evolution is more gradual or even constant. A
second sequence is located at less negative \vabs\ ($\vabs \ga
-15,000$\kms) and remains almost constant around $\sim-12,000$\kms. 
This sequence corresponds to red \catwo\ absorption
components that have a blue counterpart, and the resulting
contamination biases the measurements to higher \vabs\ (see
Sect.~\ref{Sect:hvca2}). The scatter in \vabs\ 
decreases with SN phase, irrespective of decline rate, from $\sim\pm 7000$\kms\ 
at $-10$~d to $\la \pm 3000$\kms\ at
+20--30~d. The fast-decliners overlap significantly
with the other SN~Ia, and thus cannot be used to discriminate between
subluminous and overluminous objects, contrary to claims made by
\citet{Lidman:2004}. Within the $\dmm < 1.0$ sample, two objects
(SN~1990O and 1999ee) form a \vabs\ sequence at more negative
velocities, suggesting higher explosion kinetic energies. The other 8
slow-declining SN~Ia cannot be distinguished from those with $\dmm
\ge 1.0$, at all phases.

%%%%%%%%%%%%%%%%%%%%%%%%%%%%%%%%%%%%%%%%%%%%%%%%%%%%%%%%%%%%%%%%%%%
%
%    Fig. -- vabsca2red
%
%%%%%%%%%%%%%%%%%%%%%%%%%%%%%%%%%%%%%%%%%%%%%%%%%%%%%%%%%%%%%%%%%%%
%
% Figure showing Ca II absorption velocities
%

In the lower panel of Fig.~\ref{Fig:vabsca2red}, we overplot the
\vabs\ measurements for our sample of high-$z$ ($0.16<z<0.64$)
SN~Ia, again ordered according to their decline rate:
$\dmm<1.0$ ({\it stars}), and $1.0<\dmm<1.7$ ({\it squares}). The time
evolution of \vabs\ for \catwo\ \l3945 in our high-$z$ sample is
similar to that for the local SN~Ia: a steady increase from $\ga
-25,000$\kms\ at very early phases ($\la -10$~d) to $\sim-15,000$\kms\
at maximum, and a more gradual post-maximum increase. Again, the
slow-declining high-$z$ SN~Ia cannot be distinguished. 

\begin{figure}
\epsscale{1.2}
\plotone{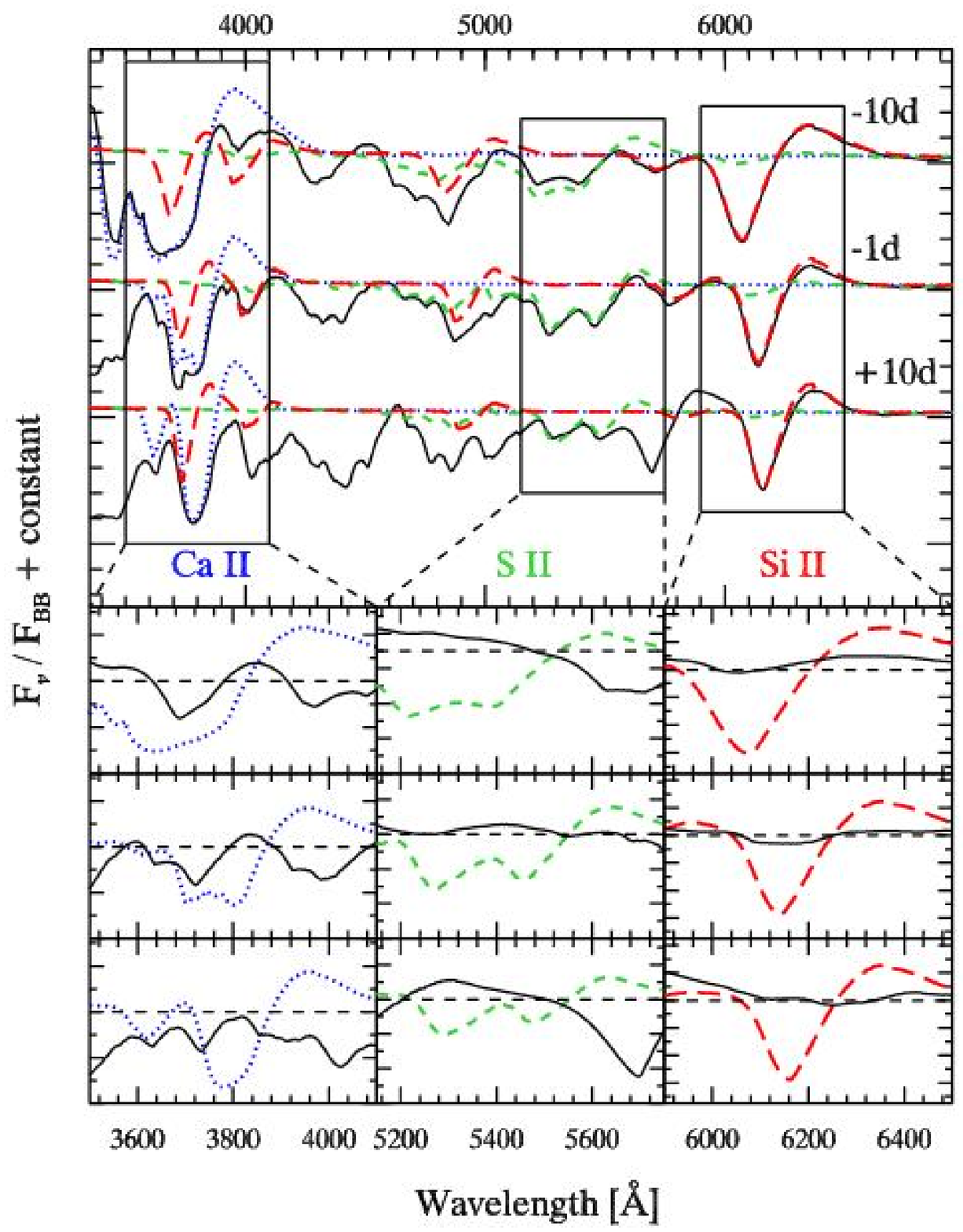}
\caption{{\it Top panel:} SYNOW  models of SN~1994D at $-$10~d, $-$1~d,
and +10~d ({\it solid line}), along with the contributions from \catwo\
({\it dotted lines}), \stwo\ ({\it dashed line}), and \sitwo\ ({\it
long-dashed line}). The fluxes (per unit frequency, $F_\nu$) are normalized to the
underlying blackbody continuum ($F_{\rm BB}$). Unlike
\citet{Branch/etal:2005} we have not included (weak) contributions
from \ctwo\ in the $-$10~d spectrum.
{\it Lower panels:} Close-up of the \catwo\ \l3945 ({\it left}),
\stwo\ $\lambda\lambda$5454, 5640 ({\it middle}), and \sitwo\ \l6355
({\it right}) features. The linestyle coding is the same as for the
top panel, except that now the solid line corresponds to the
contribution from all {\it other} ions. The fluxes are again in
$F_\nu$, and normalized to $F_{\rm BB}$ ({\it dashed line}).
See the electronic edition of \aj\ for a color version of this figure.
\label{Fig:94d_components}}
\end{figure}

Because this \catwo\ feature is a few hundred {\AA}ngstroms wide, it likely
overlaps with other lines. To illustrate and assess the magnitude of
such a line overlap, we show in the top panel of
Fig.~\ref{Fig:94d_components} synthetic SYNOW spectra for SN~1994D at
$-10$~d, $-1$~d, and $+10$~d from $B$-band maximum
(\citealt{Branch/etal:2005}; {\it solid line}), as well as the relative
contribution from \catwo\ ({\it dotted line}), \stwo\ ({\it dashed line}), and
\sitwo\ ({\it long-dashed line}). (All spectra are normalized to the adopted
blackbody continuum energy distribution.) We see that the strong \catwo\
\l3945 absorption feature (including the blue absorption, see
Sect.~\ref{Sect:hvca2}) is contaminated at all phases, predominantly
by \sitwo\ \l3858 ($gf$-weighted mean rest-frame wavelength). However, it
is only around and past maximum that the \vabs\ measurement is
affected by the \sitwo\ \l3858 absorption: at maximum (+10~d), \vabs\
is biased to less (more) negative values. Despite this corrupting
effect, the \catwo\ \l3945 line is the major contributor to the wide
absorption trough seen at $\sim3750$~\AA, for all phases $\la 2$ weeks
from $B$-band maximum.

%%%%%%%%%%%%%%%%%%%%%%%%%%%%%%%%%%%%%%%%%%%%%%%%%%%%%%%%%%%%%%%%%%%
%
%    Fig. -- 94d_components
%
%%%%%%%%%%%%%%%%%%%%%%%%%%%%%%%%%%%%%%%%%%%%%%%%%%%%%%%%%%%%%%%%%%%
%
%
%

\begin{figure}	
\epsscale{1.1}
\plotone{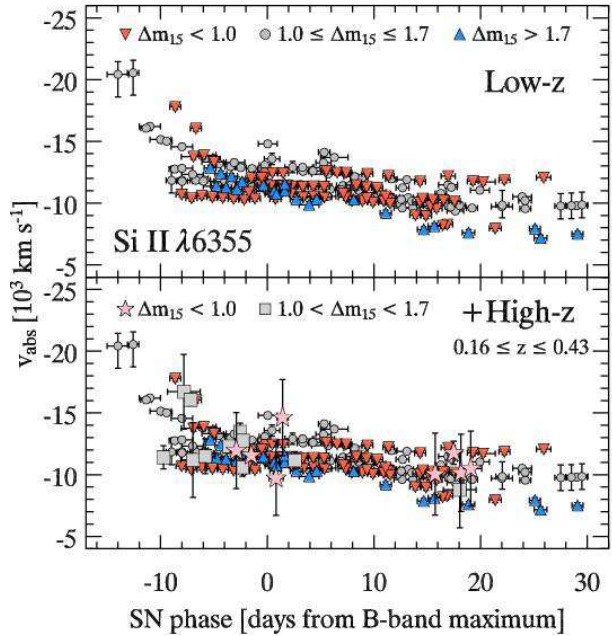}
\caption{{\it Upper panel:} Absorption velocities for \sitwo\ \l6355 
in local SN~Ia, for three different \dmm\ ranges. {\it Lower panel:}
The high-$z$ data are overplotted.
See the electronic edition of \aj\ for a color version of this figure.
\label{Fig:vabssi2}}
\end{figure}

\begin{figure*}	
\begin{center}
\includegraphics[width=10cm, angle=90]{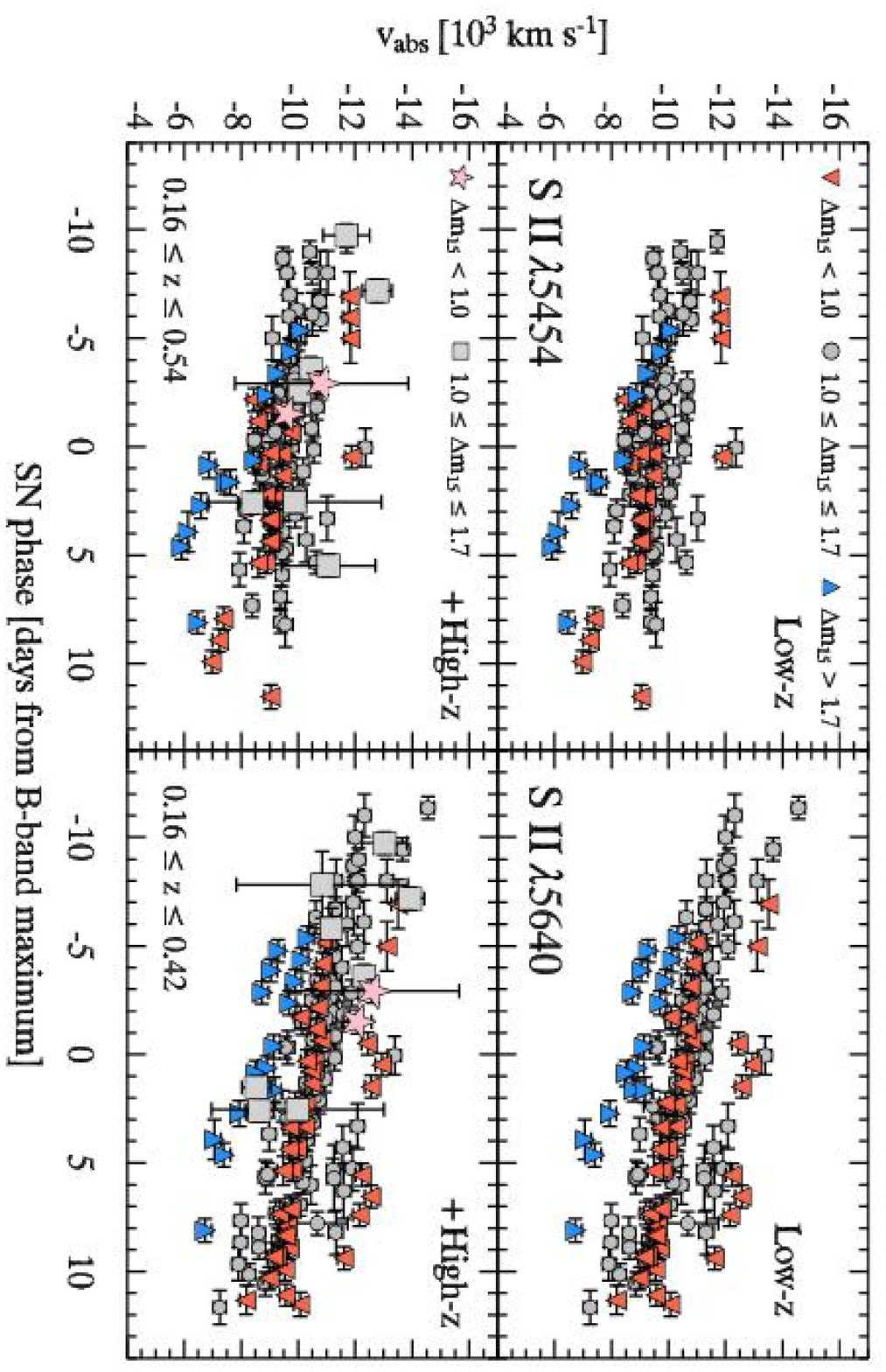}
\caption{{\it Upper panels:} Absorption velocities for \stwo\ \l5454
({\it left}) and \stwo\ \l5640 ({\it right}) in local SN~Ia, for three different
\dmm\ ranges. {\it Lower panels:} The high-$z$ data are overplotted.
See the electronic edition of \aj\ for a color version of this figure.
\label{Fig:vabss2}}
\end{center}
\end{figure*}

In Fig.~\ref{Fig:vabssi2} (top panel), we reproduce
Fig.~\ref{Fig:vabsca2red} for \sitwo\ \l6355, showing in the top panel the
\vabs\ measurements for the local SN~Ia sample. The \vabs\ evolution
for this feature is comparable to that for \catwo\ \l3945, though values
are at all phases less negative, by $\sim5000$\kms\ 
(see also Fig.~\ref{Fig:94d_components} and Sect.~\ref{Sect:vabscmf}).
The fast-declining SN~Ia form, on average, a sequence of less
negative \vabs, at post-maximum phases, but this 
sequence separates only at $t \ga +20$~d from the $1.0 \le \dmm \le
1.7$ objects. The higher scatter in the slow-declining SN~Ia causes
an overlap with the fast-declining ones at all phases. We are lacking
data at $t \ga +20$~d to make a clear distinction between the slowest
and fastest decliners of our sample. At these late phases, however,
the optical spectra of SN~Ia are dominated by lines of iron-group
elements (mainly \cotwo\ and \fetwo), and the \sitwo\ \l6355 feature suffers
from increasing line blending. Note that the \sitwo\ \l6355 absorption
profile is sometimes unusually flat and extended (see, e.g., SN~1990N
at $-14$~d and $-13$~d with $\vabs \sim-20,000$\kms; also see
SN~2001el in \citealt{Mattila/etal:2005}), perhaps due to
contamination from \ctwo\ \l6580 forming in a high-velocity shell
(\citealt{Fisher/etal:1997}; see also \citealt{Mazzali:2001}). The
high-$z$ sample (Fig.~\ref{Fig:vabss2}, lower panel) reveals similar
properties. Note that \sitwo\ \l6355 falls outside the optical spectral
range for $z \ga 0.4$; the highest redshift at which we were able to
measure \vabs\ for this feature was $z=0.428$.

%%%%%%%%%%%%%%%%%%%%%%%%%%%%%%%%%%%%%%%%%%%%%%%%%%%%%%%%%%%%%%%%%%%
%
%    Fig. -- vabssi2
%
%%%%%%%%%%%%%%%%%%%%%%%%%%%%%%%%%%%%%%%%%%%%%%%%%%%%%%%%%%%%%%%%%%%
%
% Figure showing Si II absorption velocities
%

We now turn to the weakest lines in our study, the \stwo\ \l\l5454, 5640
features, for which we show the time-evolution of \vabs\ in
Fig.~\ref{Fig:vabss2}. Compared to \catwo\ \l3945 and \sitwo\ \l6355,
\vabs\ values for the \stwo\ lines are less negative at all phases (always
greater than $-13,000$\kms\ for \l5454 and $-15,000$\kms\ for \l5640), with a
nearly constant and smaller increase with phase. These
optically thinner lines are increasingly contaminated by those of
iron-group elements at $t \gtrsim 2$ weeks, becoming unnoticeable at
later phases -- this explains the lack of data at late phases in
Fig.~\ref{Fig:vabss2}. Three of the $1.0 \le \dmm \le 1.7$ points
for \stwo\ \l5640 at $t \sim-10$~d, associated with SN~2002bo
\citep{Benetti/etal:2004}, lie at more negative velocities ($< -13,000$\kms)
than the bulk of our sample. Several points at phases between maximum
and +10~d, associated with SN~1994M \citep{Gomez/Lopez:1996}, also
have more negative velocities than the rest of our sample data. This may
result from line blending, although our SYNOW investigation (see
Fig.~\ref{Fig:94d_components}, middle panels) suggests this overlap
to be weak or absent.

%%%%%%%%%%%%%%%%%%%%%%%%%%%%%%%%%%%%%%%%%%%%%%%%%%%%%%%%%%%%%%%%%%%
%
%    Fig. -- vabss2
%
%%%%%%%%%%%%%%%%%%%%%%%%%%%%%%%%%%%%%%%%%%%%%%%%%%%%%%%%%%%%%%%%%%%
%
% Figure showing S II 5454 and S II 5640 absorption velocities
%

Alternatively, the \stwo\ \vabs\ measurements might be influenced by the
overlap between the 13 individual \stwo\ features in the range
5300--5700~\AA\ (Kurucz \& Bell 1995). For example, distinct intrinsic
excitation temperatures and formation mechanisms for one transition
could cause its optical depth to vary differently from that of others
and thus modulate, at selected phases, the observed location of
maximum absorption in the total profile. Note that the \catwo\ \l3945 and
\sitwo\ \l6355 features are the result of transitions corresponding to a
single multiplet, and will not be affected by this issue. 

Interestingly, fast decliners at positive phase show the least negative \stwo\
\l5454--\l5640 \vabs\ values with a lower limit of $\sim-8000$\kms,
combined with the most pronounced \vabs\ gradient with phase amongst the
different \dmm\ subgroups. Provided the phase is known accurately,
these two \stwo\ features can be used to discriminate between fast- and
slow-decliners at post-maximum phases.

\subsubsection{Interpretation of \vabs\
measurements\label{Sect:vabscmf}} 

We now investigate the causes of the variations in line profile
shapes and $\vabs$ values for our optical diagnostics. We base our
discussion on synthetic line profiles computed  with CMFGEN
\citep{Hillier/Miller:1998}, a steady-state, one-dimensional, 
non-local thermodynamic equilibrium (non-LTE) model
atmosphere code that solves the radiative transfer equation in the
comoving frame, subject to the constraints of radiative and
statistical equilibria. Because CMFGEN is at present not strictly
adequate for SN~Ia conditions (no chemical stratification; no
$\gamma$-ray energy deposition; neglect of relativistic effects apart from
first-order Doppler corrections; see \citealt{Dessart/Hillier:2005a} for
details), these results are merely illustrative; nonetheless, they
provide a new insight into the sites of optical-line and continuum
formation, corresponding in this example to a low-luminosity SN~Ia
(``SN 1991bg-like'') near maximum light.

%%%%%%%%%%%%%%%%%%%%%%%%%%%%%%%%%%%%%%%%%%%%%%%%%%%%%%%%%%%%%%%%%%%
%
%    Fig. -- cmfgen 
%
%%%%%%%%%%%%%%%%%%%%%%%%%%%%%%%%%%%%%%%%%%%%%%%%%%%%%%%%%%%%%%%%%%%
%
% Figure showing line profiles for a CMFGEN model of 1991bg 
%
\begin{figure*}	
\begin{center}
\includegraphics[width=10cm, angle=90]{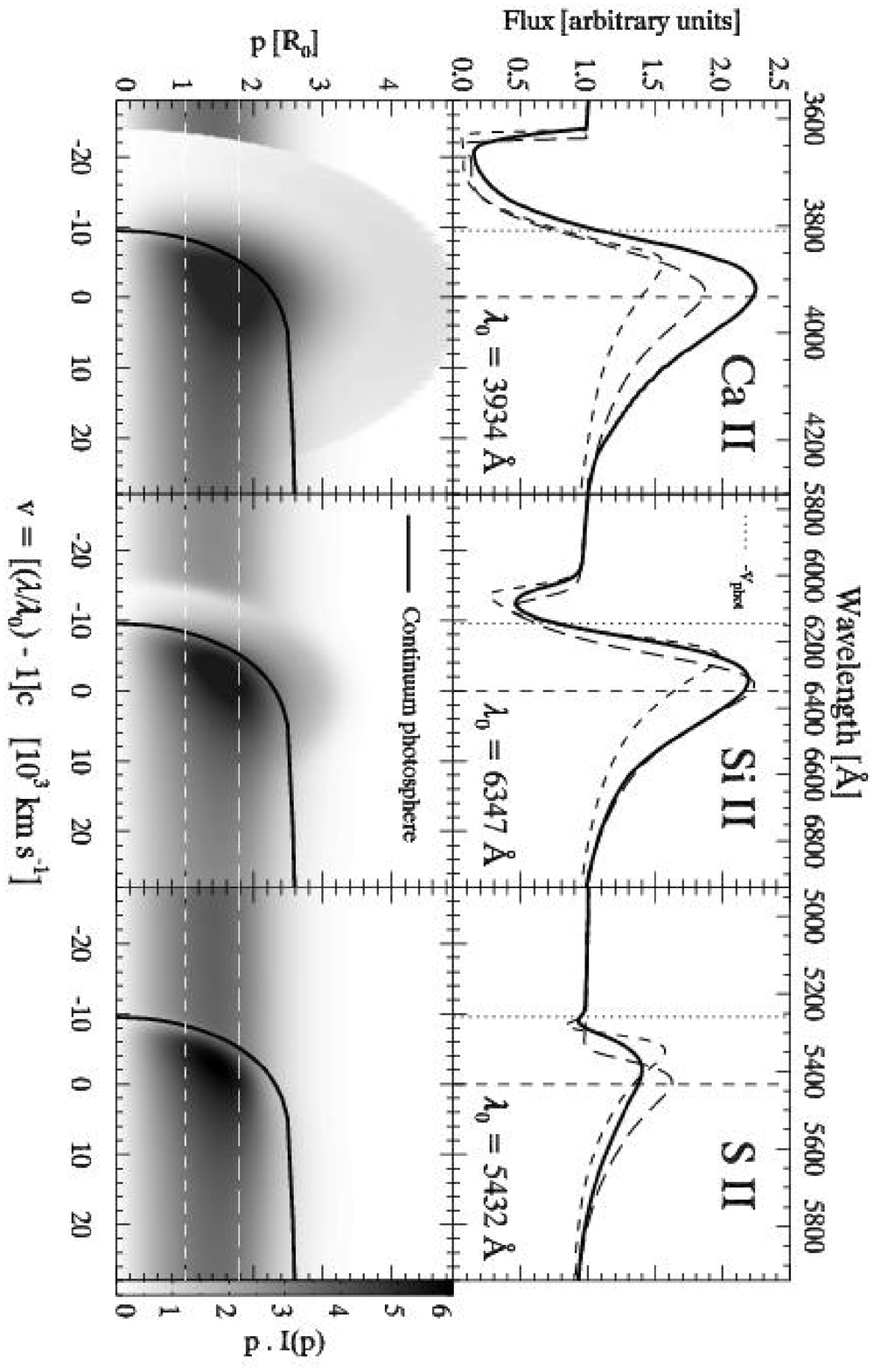}
\caption{P-Cygni profiles of \catwo\ \l3934, \stwo\ \l5432, and \sitwo\ \l6347
in a CMFGEN model of a low-luminosity SN~Ia near maximum brightness, with 
density exponent $n=7$, revealing the sites at the origin of synthetic
line profile flux, and the resultant blueshift of the P-Cygni profile
emission peak. {\it Lower panels:} Grayscale image of the quantity $p
\cdot I(p)$ as a function of $p$ and classical Doppler velocity $v =
[(\lambda/\lambda_0) - 1]c$, where $p$ is the impact parameter
and $I(p)$ the emergent specific intensity along $p$ (at $v$). 
$R_0$ is the base radius of the CMFGEN radial grid where the continuum optical
depth $\tau_{\rm cont} \approx 50$ -- a photosphere thus exists
in this model configuration, corresponding here to a velocity of
9550 \kms. $\lambda_0$ is the  
rest wavelength of the transition and $c$ is the speed of light in
vacuum. The overplotted thick black curve gives the line-of-sight 
velocity location where the integrated continuum optical depth at
5500~\AA, along $z$ and at a given $p$, equals 2/3 
(Note that the photospheric velocity quoted above -- not projected -- 
is found at the point on this curve with $p=0$, which also corresponds to a depth 
$z = 1.91 R_0 = v_{\rm phot}/v_0$; see Table~\ref{Table:cmfgenparams}).
In the ($p,z$) plane, this curve has a similar shape but more circular
for negative $z$, as shown for a synthetic H$\alpha$ profile for a Type II 
SN in Fig.~10 of \citet{Dessart/Hillier:2005b}.
The dotted white lines are for $p$ = [1,1.8]. {\it Upper panels:} 
({\it solid curve}) Line
profile flux obtained by summing $p \cdot I(p)$ over the range of
$p$; ({\it broken curves}) velocity profile of $p \cdot I(p)$ for 
two $p-$rays, at $p$ = [1,1.8].
The vertical dotted line corresponds to the (continuum) photospheric
velocity. The profiles have been normalized to unity  
at the inferior boundary of the plotted velocity range, where the line
optical depth is zero. See
\citet{Dessart/Hillier:2005a,Dessart/Hillier:2005b} for a detailed and
pedagogical explanation of these plots in the context of Type II
supernovae.
\label{Fig:cmfgen}}
\end{center}
\end{figure*}

The SN~Ia conditions are epitomized here by the absence of hydrogen
and helium in the outflow, and thus the dominance of metal species. 
Their mass fractions, alongside basic model parameters, are listed
in Table~\ref{Table:cmfgenparams}. Note that the continuum optical
depth at the base radius $R_0$ is $\sim$50.

%%%%%%%%%%%%%%%%%%%%%%%%%%%%%%%%%%%%%%%%%%%%%%%%%%%%%%%%%%%%%%%%%%%
%
%    Table -- CMFGEN model parameters
%
%%%%%%%%%%%%%%%%%%%%%%%%%%%%%%%%%%%%%%%%%%%%%%%%%%%%%%%%%%%%%%%%%%%
%
% 
%
\begin{center}
\begin{deluxetable}{l l}
\tablenum{5}
%\tabletypesize{\scriptsize}
\tablewidth{0pt}
\tablecaption{CMFGEN model parameters
\label{Table:cmfgenparams}}  
\tablehead{
\colhead{Model parameter} & 
\colhead{Parameter value}}
\startdata                  
Radius, $R_0$         & $4 \times 10^{14}$~cm ($\sim5750R_\sun$)         \\
Velocity at $R_0$, $v_0$     & 5000\kms \\
Luminosity, $L_0$     & $8 \times 10^8 L_\sun$             \\
Density, $\rho_0$     & $2 \times 10^{-11}$~g~cm$^{-3}$ \\
Density gradient, $n$ & 7 [in $\rho(r) = \rho_0 (R_0/r)^n$]     \\
Turbulent velocity, $v_{\rm turb}$ & 90\kms                   \\
\hline
Element            & Mass fraction                            \\
\hline
C                  & 0.12                                     \\
O                  & 0.63                                     \\
Mg                 & 0.1                                      \\
Si                 & 0.1                                      \\
S                  & 0.05                                     \\
Ca                 & 0.0001                                   \\
Fe                 & 0.0014                                   \\
Ni                 & 0.001                                    \\
\enddata
\end{deluxetable}
\end{center}

We show in Fig.~\ref{Fig:cmfgen} the synthetic line profiles for \catwo\
\l3934 ({\it left}), \sitwo\ \l6347 ({\it center}), and \stwo\ \l5432 ({\it
right}), computed 
under such model assumptions. At the bottom of each panel, and
following \citet{Dessart/Hillier:2005a,Dessart/Hillier:2005b}, we show
grayscale images in the $(v,p)$ plane of the flux-like quantity $p
\cdot I(p)$, where $v = [(\lambda/\lambda_0) - 1]c$ is the
classical Doppler velocity, $p$ is the impact parameter in units of
$R_0$, and $I(p)$ is the specific intensity at $p$. 
The photosphere, corresponding to $\tau_{\rm cont} = 2/3$ at 5500~\AA,
is shown as a solid thick line. The dominance of the (gray)
electron-scattering opacity makes the corresponding radius essentially
wavelength independent over the range considered here; see, e.g.,
Sect.~4.2 and Figs.~7--9 of \citet{Dessart/Hillier:2005b}. 
The sum over $p$ of the quantity $p \cdot I(p)$ at
$v$ corresponds to the total line flux at $v$, shown at the top of
each panel (solid line). Note that for each line, we select a single
transition to avoid the corrupting effect of line overlap, stemming
from other transitions of the same or different species.

Let us first focus on the absorption trough, controlling the resulting
$\vabs$, of such synthetic line profiles. For \catwo\ \l3934, we see that
the trough is nearly saturated with essentially no residual flux down
to $\sim-22,000$\kms, while \sitwo\ \l6347 shows a maximum absorption
at a less negative velocity of $-$12,000\kms; \stwo\ \l5432 is the weakest line
of all three, with a very modest absorption and extent, located at
$\sim-9500$\kms. 

We thus reproduce here the general trend shown in
Figs.~\ref{Fig:vabsca2red}-\ref{Fig:vabss2} and presented in the
previous section: absorption velocities for \catwo\ \l3945 are 
more negative by several 1000\kms\ at any given phase than those for \stwo\
\l\l5454,5640  and \sitwo\ \l6355, because it remains optically thick out
to larger radii (i.e., lower densities and higher expansion
velocities).  Indeed, this feature is the result of a blend of \catwo\ K 
(3933.66~\AA) and H (3968.47~\AA) transitions, both corresponding to
the same 4s$^2$S--4p$^2$P$^0$ multiplet, linking the ground state and
low-lying upper levels (just 3~eV above the ground state). Despite the
low $\log(gf)$-value of the transition and the considerably lower calcium
abundance compared to silicon and sulfur in our model  (by a factor
of 1000 and 500, respectively), the high \catwo\ ground-state population in this
parameter space translates into a very large line optical depth
($\tau_{\rm line} \propto \kappa_{\rm line} \rho_{\rm ion}$). The
sample \sitwo\ and \stwo\ lines result, however, from higher-level
transitions, less populated, which translate into systematically lower
optical depths and less negative $\vabs$-values, the more so for the \stwo\
lines. Also, at a given phase, the maximum absorption  
is further to the blue in \stwo\ \l5640 than in \stwo\ \l5454, which
likely results from differences in the atomic properties of each
transition.

Despite the assumed {\it homogeneity and smooth density distribution}
of the CMFGEN model, a scatter in $\vabs$ between line diagnostics is
not only present but also large and comparable to the observed
scatter; inferring 
the presence, at a given phase, of chemical stratification in the SN
outflow thus requires careful analysis, with a detailed and accurate
account of all line optical depth effects
\citep{Stehle/etal:2005}. This argues for caution in
the interpretation of $\vabs$ measurements, since one sees the numerous
competing effects arising from differences in the atomic-transition
properties, chemical abundances, density, and velocity distribution;
additional corrupting effects such as line overlap are discussed later
in this paper.

In Fig.~\ref{Fig:cmfgen}, we also see that $|\vabs|$ underestimates the
velocity of the photosphere for the weak,
optically thinner \stwo\ \l5432 line. The physical origin of this effect
is given in \citet{Dessart/Hillier:2005b}, and stems from the steep
density gradient in supernova atmospheres [$\rho(r) \propto r^{-n}$]:
iso-velocity curves are at constant $v$ (depth $z$ in the $(p,z)$
plane) but the density varies as $1/r^n$. Thus, at fixed $z$, the
density drops fast for increasing $p$, at the same time reducing the
probability of line scattering and/or absorption. In practice, the
location of maximum absorption along a $p-$ray shifts to larger depths
($z$ closer to zero) for increasing $p$, showing overall the same
curvature as seen for the photosphere (see overplotted
curve). Along a given $p-$ray, the location of
maximum absorption is always exterior to the 
photosphere along that ray (the line opacity comes on top of the
default continuum opacity), but shifts toward line center  for
increasing $p$. As a consequence, the total line 
profile, which results from the contribution at all impact parameters,
shows a maximum absorption at a velocity \vabs, the magnitude of
which can be higher
{\it or} lower than the photospheric value $v_{\rm
phot}$. This offset between $|\vabs|$ and $v_{\rm phot}$ is determined
primarily by the magnitude of the line optical depth. As we move from
\catwo\ \l3934 to \sitwo\ \l6347 and \stwo\ \l5432, the line optical depth
decreases and the corresponding absorption velocity is closer to zero. 

Note that the comparison with the photospheric velocity of the SN~Ia
outflow is not necessarily meaningful. First, electron scattering,
which provides the dominant source of continuum opacity in ionized
hydrogen-free and helium-free outflows, corresponds to a small mass
absorption coefficient due to the high mean molecular weight of the
gas ($\kappa_e \la$ 0.05 cm$^2$~g$^{-1}$, a factor at least ten
times smaller than in hot-star outflows). Second, the outflow density 
decreases with the cube of the time, following the homologous expansion
of the SN~Ia, so that the outflow becomes optically thin in the
continuum after about 10 days past explosion; the concept of a
photosphere then becomes meaningless. The alternative definition (not
adopted here) of the photosphere as the location where the total
(line and continuum) inward integrated optical depth is 2/3
changes this conception somewhat: the ubiquitous presence of lines
makes the photospheric radius (and velocity) highly dependent on 
wavelength, and thus non-unique and ambiguous  
\citep[see][]{Hoeflich/etal:1993, Spyromilio/Pinto/Eastman:1994, 
Hoeflich:1995, Pinto/Eastman:2000}.

%%%%%%%%%%%%%%%%%%%%%%%%%%%%%%%%%%%%%%%%%%%%%%%%%%%%%%%%%%%%%%%%%%%
%%
%%   Emission-peak velocities
%%
%%%%%%%%%%%%%%%%%%%%%%%%%%%%%%%%%%%%%%%%%%%%%%%%%%%%%%%%%%%%%%%%%%%

\subsection{Emission-Peak Velocities\label{Sect:vpeak}}

Large negative \vpeak\ values are common in optical line profiles of
Type II SN spectra, explained for the first time by Dessart \& Hillier
(2005a); the root cause is the strong SN outflow density gradient, and
because such a property is common to both SN~Ia and SN~II, such
peak-emission blueshifts are also expected in SN~Ia line
profiles. Using the same approach as before for \vabs, we present in
Sect.~\ref{Sect:vpeakobs}, for the first time, a census of \vpeak\
measurements, using our large sample of local and  high-$z$ SN~Ia
spectra. We then comment on these results in
Sect.~\ref{Sect:blueemission}, using the CMFGEN model presented in the
previous section.

\subsubsection{Presentation of \vpeak\ Measurements}
\label{Sect:vpeakobs}

\begin{figure}	
\epsscale{1.1}
\plotone{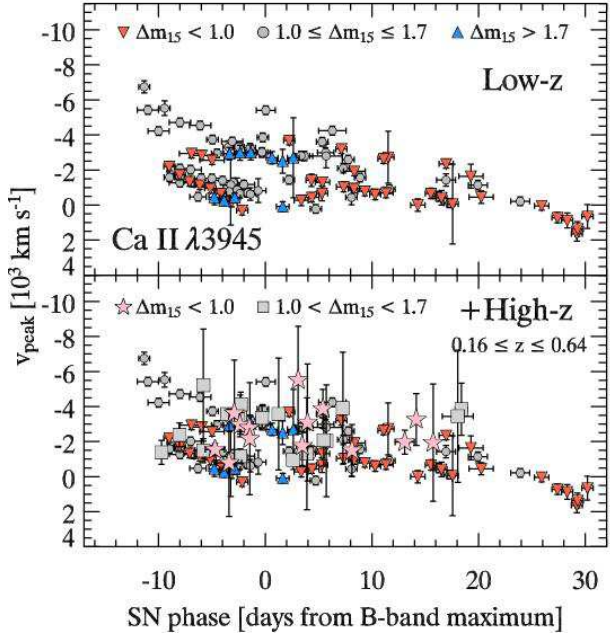}
\caption{{\it Upper panel:} Emission-peak velocities for \catwo\ \l3945 
in local SN~Ia, for three different \dmm\ ranges. {\it Lower panel:}
The high-$z$ data are overplotted. 
See the electronic edition of \aj\ for a color version of this figure.
\label{Fig:vpeakca2}}
\end{figure}

Keeping the same structure as for previous figures displaying \vabs\ 
measurements, we show in Fig.~\ref{Fig:vpeakca2} the phase-dependent
\vpeak\ measurements for the \catwo\ \l3945 feature in local SN~Ia
(upper panel) and all sampled SN~Ia (lower panel), separating objects
according to their decline-rate parameter, \dmm. As for SN~II,
all \vpeak\ measurements for \catwo\ \l3945 at $\la +20$~d are {\it
negative}; that is, the peak emission is blueshifted with respect to the
rest wavelength of the line, increasing with phase from $\sim-6000$\kms\ 
(at $-$10~d) up to $-1000$\kms\ (+20~d). The scatter is,
however, significant. Moreover, at any phase, near-zero values are
found. As previously for \vabs\ measurements, \sitwo\ \l3858 modifies the
\catwo\ intrinsic line profile shape  but now introduces, at all phases,
a blueshift of the emission peak of the 3945~\AA\ feature
(Fig.~\ref{Fig:94d_components}, left panels), likely influencing the
scatter and the magnitude of \vpeak\ values (see next section). A few
points at $\ga +25$~d (corresponding to the spectroscopically
peculiar SN~1999aa; \citealt{Li/etal:2001a}) show a
counterexample to the above trend, with $\vpeak > 0$. At these
phases, however, \catwo\ \l3945 is increasingly contaminated from lines
of iron-group elements, and the measurements of \vpeak\ at these
phases are highly uncertain. Note
that \vpeak\ measurements for the high-$z$\ sample are consistent with
the trend in the local sample, both qualitatively and quantitatively
(Fig.~\ref{Fig:vpeakca2}, lower panel).

%%%%%%%%%%%%%%%%%%%%%%%%%%%%%%%%%%%%%%%%%%%%%%%%%%%%%%%%%%%%%%%%%%%
%
%    Fig. -- vpeakca2
%
%%%%%%%%%%%%%%%%%%%%%%%%%%%%%%%%%%%%%%%%%%%%%%%%%%%%%%%%%%%%%%%%%%%
%
% Figure showing Ca II emission-peak velocities
%

\begin{figure}	
\epsscale{1.1}
\plotone{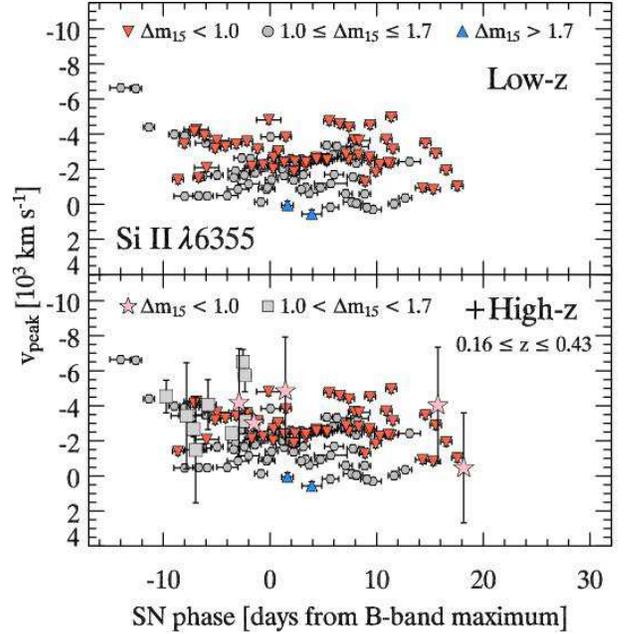}
\caption{Same as Fig.~\ref{Fig:vpeakca2} for \sitwo\
\l6355. 
See the electronic edition of \aj\ for a color version of this figure.
\label{Fig:vpeaksi2}}
\end{figure}

A similar pattern is also found for \sitwo\ \l6355, both for local and
high-$z$ SN~Ia (Fig.~\ref{Fig:vpeaksi2}), although the measurement
errors for the latter are $\sim1000$\kms. Unlike for the \catwo\
feature, these measurements are free of sizeable line overlap (see
previous section), and therefore represent a genuine intrinsic
blueshift of peak emission of \sitwo\ \l6355. Note that \vpeak\ for the
slow-declining (overluminous) SN~Ia is slightly more negative (by $\sim1000$\kms), 
on average, than for the fast-declining
(underluminous) objects. Since the magnitude of the
emission-peak blueshift scales with the expansion
velocity of the ejecta (see end of the section), 
this dependence supports the idea that slow-declining SN~Ia correspond to higher
kinetic energy explosions \citep{Mazzali/etal:1998,Leibundgut:2000}. The two 
points with $\vpeak < -6000$\kms\ at $t<-10$~d correspond to SN~1990N,
for which the \vabs\ measurements at these phases are also more
negative (Fig.~\ref{Fig:vabssi2}).

%%%%%%%%%%%%%%%%%%%%%%%%%%%%%%%%%%%%%%%%%%%%%%%%%%%%%%%%%%%%%%%%%%%
%
%    Fig. -- vpeaksi2
%
%%%%%%%%%%%%%%%%%%%%%%%%%%%%%%%%%%%%%%%%%%%%%%%%%%%%%%%%%%%%%%%%%%%
%
% Figure showing Si II emission-peak velocities
%

\begin{figure*}	
%\plotone{../vpeaks2.ps}
\begin{center}
\includegraphics[width=10cm, angle=90]{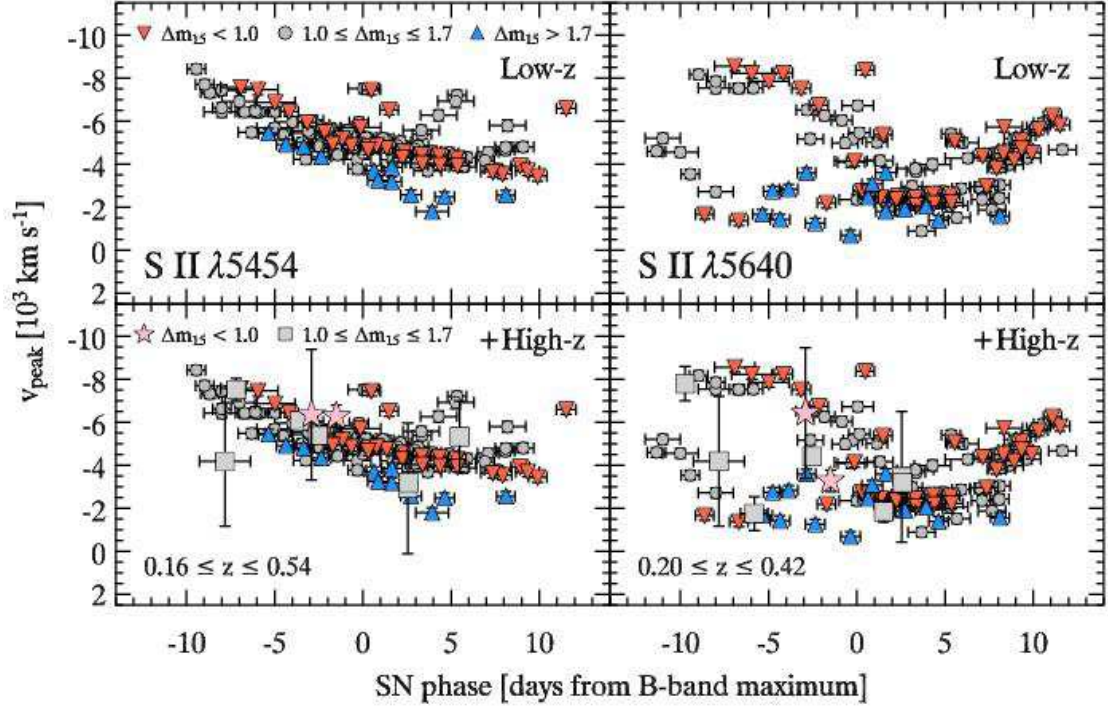}
\caption{{\it Upper panels:} Emission-peak velocities for \stwo\ \l5454 
({\it left}) and \stwo\ \l5640 ({\it right}) in local SN~Ia, for three different
\dmm\ ranges. {\it Lower panels:} The high-$z$ data are
overplotted. 
See the electronic edition of \aj\ for a color version of this figure.
\label{Fig:vpeaks2}} 
\end{center}
\end{figure*}

We finally turn in Fig.~\ref{Fig:vpeaks2} to the \vpeak\ measurements
for the \stwo\ \l5454, 5640 features. We first start with the right panels,
which show that the \stwo\ \l5640 data points, both for the local and
high-$z$ samples, follow a similar pattern of increasing values with
phase as for the \catwo\ and \sitwo\ features, with velocity shifts always
negative but now reaching down to $-8~500$\kms\ at $\sim-$10~d. We
associate the large scatter of data points to contamination by \sitwo\
\l5972 at pre-maximum phases, and \naone~D \l5892 at post-maximum
phases. Note that for the high-$z$ sample, an uncertainty of about
3000\kms\ is introduced when the SN redshift is determined {\it 
via} cross-correlation with local SN~Ia spectral templates (see Table
\ref{Table:errors}). For \stwo\ \l5454 (left panel), we find, on
average, a steady increase of \vpeak\ with phase, from $-$8~500\kms\
at $-$10~d to $-$4000\kms\ at +10~d, but with a clear
dichotomy according to \dmm\ parameter: fast-decliners show
systematically faster-increasing and less-negative values,
related to the 
modest expansion velocity of their outflows and the larger (comoving)
recession velocity of the photo-emitting layers in the SN ejecta. The
low scatter of data points, due to the absence of sizeable
line overlap, makes this distinction clear and suggests that, as
before from \vabs\ measurements, the blueshift of peak emission at
post-maximum phases can now also be used to isolate fast-declining
SN~Ia, provided the SN phase is accurately known.

%%%%%%%%%%%%%%%%%%%%%%%%%%%%%%%%%%%%%%%%%%%%%%%%%%%%%%%%%%%%%%%%%%%
%
%    Fig. -- vpeaks2
%
%%%%%%%%%%%%%%%%%%%%%%%%%%%%%%%%%%%%%%%%%%%%%%%%%%%%%%%%%%%%%%%%%%%
%
% Figure showing S II 5454 and S II 5640 emission-peak velocities
%

\begin{figure}	
\epsscale{1.1}
\plotone{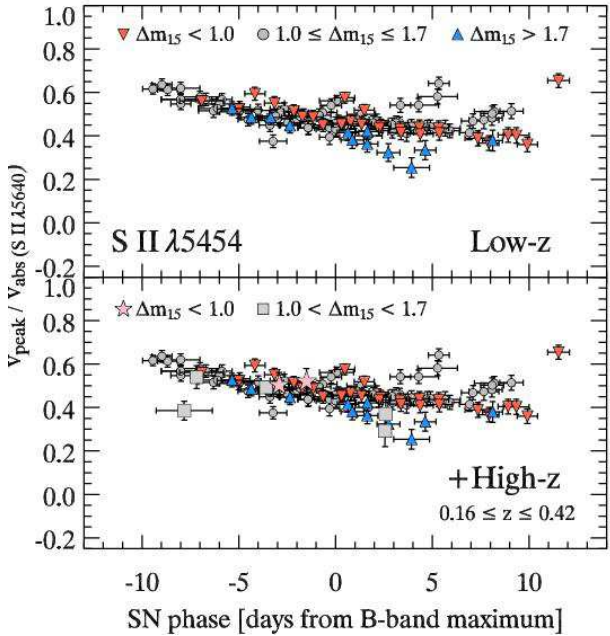}
\caption{{\it Upper panel:} Emission-peak velocities for \stwo\ \l5454
normalized to the \stwo\ \l 5640 absorption velocities. {\it Lower
panel:} The high-$z$ data are overplotted.
See the electronic edition of \aj\ for a color version of this figure. 
\label{Fig:vpeaks2blue_norm2s2red}}
\end{figure}

To conclude this section on the \vpeak\ measurements, we
show in Fig.~\ref{Fig:vpeaks2blue_norm2s2red} the values for \stwo\
\l5454, but this time normalized to $v_{\rm abs,5640}$ at the same
phase, a quantity that closely matches, at early times, the photospheric velocity of
the flow (Fig.~\ref{Fig:cmfgen}). For both local and high-$z$ SN~Ia,
this ratio covers the 0.2--0.7 range, and thus represents a
significant shift of a line profile, comparable to the shift
identified in the measurement of the absorption velocity. The
time evolution of \vpeak/\vabs\ has a flatter slope than the
corresponding \vpeak\ sequence of Fig.~\ref{Fig:vpeaks2} (left
panels), suggesting that the magnitude of emission blueshift is a good
tracer of the expansion velocity.

%%%%%%%%%%%%%%%%%%%%%%%%%%%%%%%%%%%%%%%%%%%%%%%%%%%%%%%%%%%%%%%%%%%
%
%    Fig. -- vpeaks2blue_norm2s2red
%
%%%%%%%%%%%%%%%%%%%%%%%%%%%%%%%%%%%%%%%%%%%%%%%%%%%%%%%%%%%%%%%%%%%
%
% Figure showing S II 5454 emission-peak velocities normalized to
% vabs(s2 5454)
%

%%%%%%%%%%%%%%%%%%%%%%%%%%%%%%%%%%%%%%%%%%%%%%%%%%%%%%%%%%%%%%%%%%%
%%%%%%%%%%%%%%%%%%%%%%%%%%%%%%%%%%%%%%%%%%%%%%%%%%%%%%%%%%%%%%%%%%%

\subsubsection{Physical Origin of the Emission-Peak
Blueshift\label{Sect:blueemission}}

To investigate the origin of the observed blueshift of peak emission,
let us go back to the left panel of Fig.~\ref{Fig:cmfgen} and study
the sites of emission in the \catwo\ \l3934 line. As in the standard
cartoon of P-Cygni profile formation, one can view a significant amount
of flux arising from the side lobes, corresponding to regions with $p
> p_{\rm lim} \approx 3 R_0$ -- this defines the spatial extent of the
(continuum) photodisk. Despite the weaker emission at such distances from the
``photosphere,'' the total contribution is quite bigger because it
involves a larger and optically-thinner emitting volume. Contrary to such
a heuristic P-Cygni profile formation, a significant amount of
emission arises also from the region with $p < p_{\rm lim}$; this more
restricted volume is however affected by continuum optical depth effects
since it resides partly within the photosphere (represented
by the solid dark line). Although this latter emission source appears
in the blue side of the profile, the larger contribution from the
lobes leads to a relatively symmetric emission peak.

Moving to the optically thinner \sitwo\ \l6355 line (middle
panel), we now see that the relative (symmetric) emission contribution
from the lobe is lower compared to that arising from within the limbs
of the photodisk, leading to a more pronounced blueshift of peak
emission. This situation becomes even more extreme in the case of \stwo\
\l5432 (right panel), whereby no side-lobe emission is present: the
resulting P-Cygni profile shows a strongly blueshifted centroid,
corresponding to a sizeable fraction of the velocity at maximum
absorption. For this very optically thin line, continuum optical depth
effects are severe.

These synthetic line profiles are computed by accounting solely for
the opacity of the chosen line (plus all sources of continuum
opacity); thus, they lift any ambiguity brought upon by line
overlap. In view of these results, the large and scattered observed
(negative) \vpeak\ measurements for \catwo\ \l3945 are likely caused by line
overlap, the most likely candidate being \sitwo\ \l3858. However, the
measured and sizeable blueshifts for the \sitwo\ and \stwo\ lines are indeed
expected theoretically; CMFGEN computations also predict the larger
blueshift velocity (more negative \vpeak) for the
optically thinner \stwo\ lines, compared to either \sitwo\ or \catwo\
diagnostics. 

Interestingly, \citet{Kasen/etal:2002} argued that such blueshifted
emission must stem from peculiar effects, for example of a non-LTE
nature, rather than from modulations of the line source function or
optical depth. It now seems that their combined assumptions of a sharp
photosphere, the neglect of continuum opacity, and a pure scattering
source function enforce the symmetry of P-Cygni profile emission;
here, CMFGEN demonstrates that such assumptions may be invalid for a
number of lines, particularly at epochs where the ejecta are optically
thick in the continuum.

%%%%%%%%%%%%%%%%%%%%%%%%%%%%%%%%%%%%%%%%%%%%%%%%%%%%%%%%%%%%%%%%%%%
%%
%%   Double-absorption features in \catwo \l3945
%%
%%%%%%%%%%%%%%%%%%%%%%%%%%%%%%%%%%%%%%%%%%%%%%%%%%%%%%%%%%%%%%%%%%%

\subsection{Double-Absorption Features in \catwo\ \l3945\label{Sect:hvca2}}

Double-absorption features in \catwo\ are frequently observed in local
SN~Ia, usually in the near-infrared lines at \l8498, \l8542, and
\l8662 \citep[see,
e.g.,][]{Gerardy/etal:2004,Mazzali/etal:2005,Mattila/etal:2005}.  
\citet{Gerardy/etal:2004} suggest the interaction of a circumstellar
shell within the progenitor system, but \citet{Kasen/etal:2003}
proposed a departure from sphericity in the explosion, inferred from
polarization measurements of SN~2001el
\citep{Wang/etal:2003}. \citet{Gerardy/etal:2004} also discuss the
possibility of detecting such double-absorption \catwo\
absorption features in the \l3945 doublet; alternatively, such
identification could be influenced by the overlap of the bluer component
with \sitwo\ \l3858 (see Fig.~\ref{Fig:94d_components}), predicted to
dominate past $\sim$1 week before maximum
\citep{Hoeflich:1995,Hoeflich/Wheeler/Thielemann:1998,Lentz/etal:2000}. 
The association of the blue component in \catwo\
\l3945 with high-velocity \catwo\ absorption is uncertain, and we
therefore prefer referring to the {\it observed} blue/red, 
in place of the {\it interpreted} high-velocity/low-velocity components
as commonly used in the literature.

\begin{figure}	
\epsscale{1.1}
%\epsscale{.8}
\plotone{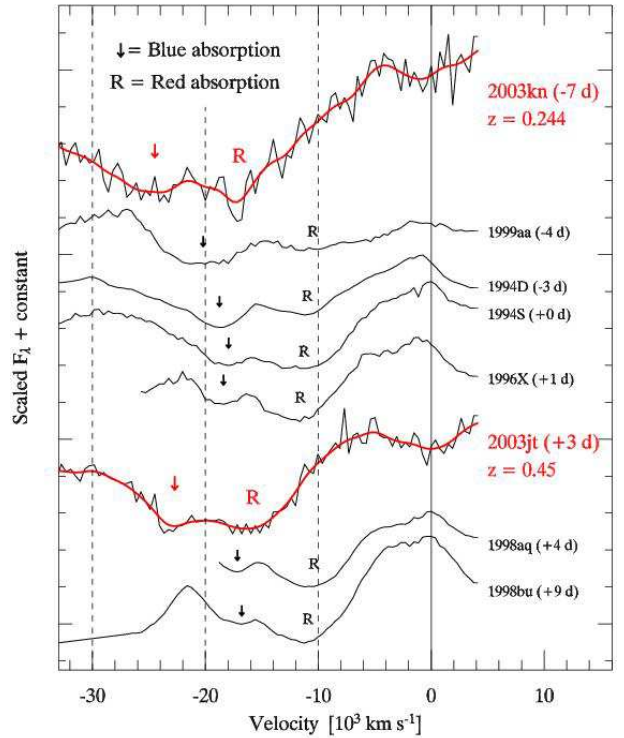}
\caption{Double-absorption features in \catwo\ \l3945. Two of our
high-$z$ SN~Ia also show a double-absorption feature. The thick line
overplotted on these spectra corresponds to their smoothed version
(see Sect.~\ref{Sect:filter}). Note the apparent $\sim-5000$\kms\ 
offset of the SN~2003jt spectrum with respect to local
SN~Ia at similar phases, probably due to an error in the SN redshift
(determined {\it via} cross-correlations with local SN~Ia spectra;
\citealt{Matheson/etal:2005}). The solid vertical line corresponds to
3945~\AA, whilst the dashed vertical lines correspond to blueshifts of
10,000, 20,000, and 30,000\kms. See the electronic edition of \aj\ for
a color version of this figure. 
\label{Fig:hvca}}
\end{figure}

Here, line-profile measurements on \catwo\ \l3945 reveal possible
double-absorption features for 6 (out of 22) local SN~Ia, and, for
the first time, for 2 (out of 34) high-$z$ SN~Ia (Table
\ref{Table:hzsn1adata}), SN~2003kn (at $-$7~d, $z=0.244$) and
SN~2003jt (at +3~d, $z=0.45$). In Fig.~\ref{Fig:hvca}, we show a
time sequence of the \catwo\ \l3945 region for a subset of the local and
high-$z$ SN~Ia samples, identifying the blue and red absorption
components, and showing the good correspondence between
profile shapes.

%%%%%%%%%%%%%%%%%%%%%%%%%%%%%%%%%%%%%%%%%%%%%%%%%%%%%%%%%%%%%%%%%%%
%
%    Fig. -- hvca
%
%%%%%%%%%%%%%%%%%%%%%%%%%%%%%%%%%%%%%%%%%%%%%%%%%%%%%%%%%%%%%%%%%%%
%
% Figure showing double-absorption Ca II 3945 in high-z SN~Ia 
%

\begin{figure}	
\epsscale{1.1}
\plotone{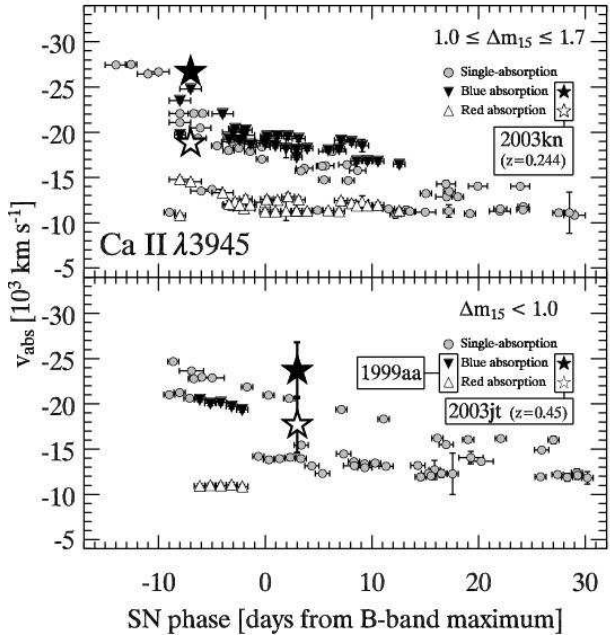}
\caption{Absorption velocities for \catwo\ \l3945 in single- and
double-absorption features (blue and red components), for objects with
$1.0 \le \dmm \le 1.7$ (upper panel) and $\dmm < 1.0$ (lower
panel). Also shown are $v_{\rm blue}$ and $v_{\rm red}$ absorption
velocities for SN~2003kn ($z = 0.244$) and SN~2003jt ($z=0.45$).} 
\label{Fig:vabs_hvca2}
\end{figure}

In Fig.~\ref{Fig:vabs_hvca2} (upper panel), we show the evolution of \vabs\ for
single- and double-absorption features in \catwo\ \l3945, selecting
objects with $1.0 \le \dmm \le 1.7$: the value of the blue (red)
component $v_{\rm blue}$ ($v_{\rm red}$) is systematically more (less)
negative than single-absorption \vabs\ values at the same phase,
perhaps caused by overlap with \sitwo\ \l3858. The blue and red data points for
the high-$z$ object (SN~2003kn) are consistent with the local SN~Ia
sample, although significantly shifted to more negative velocities and
closer together ($\sim7000$ rather than $\sim10,000$\kms); more
observations are needed to draw a firm conclusion. In the lower panel of 
Fig.~\ref{Fig:vabs_hvca2}, we show data points for objects with $\dmm
< 1.0$. Only one (SN~1999aa) out of 10 local slow-declining SN~Ia shows such a
double-absorption \catwo\ feature; one high-$z$ fast-declining SN~Ia
(SN~2003jt) out of 17 (Fig.~\ref{Fig:hvca}) also clearly displays this
feature. 

Contrary to the $1.0 \le \dmm \le 1.7$ objects, the blue
component of \catwo\ \l3945 double absorptions in SN~1999aa does not
correspond to a more negative \vabs\ than for the single
absorptions. The red component, however, lies at significantly less
negative \vabs\ ($\sim10,000$\kms\ greater than \vabs\ for the
single-absorption features at similar phases). The $\sim-5000$\kms\
vertical offset for SN~2003jt (Fig.~\ref{Fig:hvca}) is clearly seen,
although we are lacking local SN~Ia with \catwo\ \l3945 double-absorption
features at these phases. Slow decliners are often associated with  
more luminous events (though not necessarily; see Table
\ref{Table:localsn1adata}), which result from more energetic
explosions \citep{Mazzali/etal:1998,Leibundgut:2000}. It would be
worthwhile to study this spectral range in a large sample of local
slow-declining SN~Ia, to link the presence/absence of this feature
with the kinematics, and thus kinetic energy, of the explosion. 
We note the absence of \catwo\ \l3945 double absorption in our sample 
of fast-declining SN~Ia (4 SN~Ia with $\dmm > 1.7$).

%%%%%%%%%%%%%%%%%%%%%%%%%%%%%%%%%%%%%%%%%%%%%%%%%%%%%%%%%%%%%%%%%%%
%
%    Fig. -- vabs_hvca2
%
%%%%%%%%%%%%%%%%%%%%%%%%%%%%%%%%%%%%%%%%%%%%%%%%%%%%%%%%%%%%%%%%%%%
%
% Figure showing HV and LV Ca components
%

\begin{figure}	
\epsscale{1.2}
\plotone{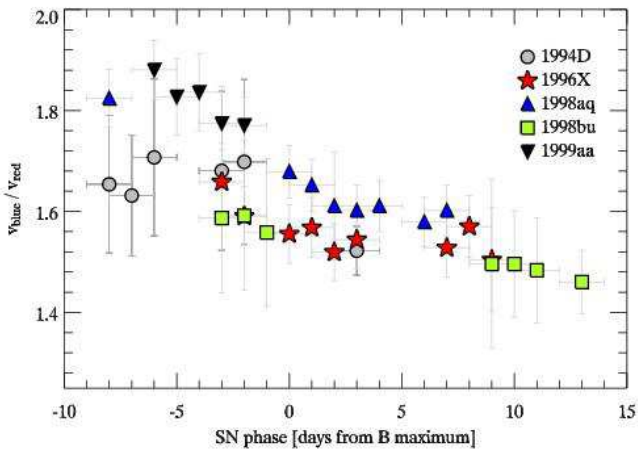}
\caption{Time evolution of the ratio of \vabs\ for the blue and red
components of the \catwo\ \l3945 absorption feature, in five local SN~Ia.
See the electronic edition of \aj\ for a color version of this figure.
\label{Fig:hvoverlv}}
\end{figure}

Finally, selecting the five local SN~Ia of our sample (SN~1994D,
1996X, 1998aq, and 1998bu) having the best temporal coverage, we show in
Fig.~\ref{Fig:hvoverlv} the evolution of the ratio $v_{\rm blue}/v_{\rm
red}$. The non-unity as well as the linear decline (except for
SN~1994D) of $v_{\rm blue}/v_{\rm red}$ over $\sim2$ weeks seems
difficult to reconcile with the presence of an inhomogeneity, expected
to leave a more transient imprint.

%%%%%%%%%%%%%%%%%%%%%%%%%%%%%%%%%%%%%%%%%%%%%%%%%%%%%%%%%%%%%%%%%%%
%
%    Fig. -- hvoverlv
%
%%%%%%%%%%%%%%%%%%%%%%%%%%%%%%%%%%%%%%%%%%%%%%%%%%%%%%%%%%%%%%%%%%%
%
% Figure showing HV/LV ratio in local SN~Ia
%

%%%%%%%%%%%%%%%%%%%%%%%%%%%%%%%%%%%%%%%%%%%%%%%%%%%%%%%%%%%%%%%%%%%
%%
%%   SECTION 4 -- DISCUSSION - CONCLUSION
%%
%%%%%%%%%%%%%%%%%%%%%%%%%%%%%%%%%%%%%%%%%%%%%%%%%%%%%%%%%%%%%%%%%%%

\section{Conclusions\label{Sect:conclusion}} 

No major systematic differences in the spectral evolution of
absorption and emission-peak velocities of several prominent lines can
be seen between local and high-$z$ SN~Ia spectra.

We present a robust measurement technique
(Sect.~\ref{Sect:filter}), which is applied to both local and
high-$z$ SN~Ia spectra. We also elaborate a reliable, if
limited, error model (errors due to blending of spectral
features cannot be reliably included). We use a spectral-smoothing
algorithm which takes
into account the Doppler broadening of SN~Ia spectral features due to
the large velocities in the ejecta, as well as the
wavelength-dependent noise affecting ground-based spectra. Our
line-profile analysis reduces the impact
of line overlap, since it relies on a smaller wavelength interval, and
allows for asymmetric line profiles. The source code is available {\it via} 
the ESSENCE web page\footnote{http://www.ctio.noao.edu/essence/}, both
as an IDL function and a Fortran script. All the results of our
measurements are displayed in Tables
\ref{Table:vabslocal}--\ref{Table:vpeakhighz}.

We find both the magnitude and time evolution of \vabs\ for SN~Ia with
different decline-rate parameters \dmm\ to be 
consistent out to $z=0.64$. As expected, strong lines have more
negative absorption velocities, and weaker lines are better tracers of the
decline-rate parameter, since they form over less extended regions. In
fact, the \stwo\ \l\l5454, 5640 features can be used as diagnostic tools to
separate fast-declining SN~Ia [$\dmm > 1.7$] from the rest, given a
reliable phase. The lack
of fast-declining SN~Ia in our high-$z$ sample prevents us from
assessing the validity of such a diagnostic at high redshifts. Most
probably the magnitude selection in ESSENCE prevents us from finding
the intrinsically less luminous, fast-declining SN~Ia at higher
redshift \citep{Krisciunas/etal:2005}. 

For the first time, we present a census of peak emission velocities,
found, up to 20~d past maximum $B$-band light, to be systematically negative.
Such a blueshifted emission peak for the three studied lines is
present in all SN~Ia of our local and high-$z$ samples,
irrespective of their decline-rate parameter. We measure \vpeak\
associated with this blueshift and find it to be a significant
fraction of \vabs\ for the \stwo\ \l5454 feature. We
show that \vpeak\ for the \stwo\ \l5454 feature can also be used to
distinguish SN~Ia with $\dmm > 1.7$ at post-maximum phases, again
given a reliable SN phase.
Using a CMFGEN model \citep{Hillier/Miller:1998}, we illustrate the
line-profile formation mechanisms in SN~Ia and show that this
blueshifted emission stems from the steep density profile
prevalent in supernova atmospheres \citep{Dessart/Hillier:2005a}.

We report the detection of double-absorption \catwo\ \l3945 
features in several local SN~Ia, and for the first time confirm its
detection in two high-$z$ SN~Ia ($z=0.244$ and $z=0.45$,
Fig.~\ref{Fig:hvca}). The association of the blue component
of this double absorption with \catwo\ is still under debate, and could
be due to contamination by \sitwo\ \l3858 \citep{Gerardy/etal:2004}.

The present investigation has not only shown the importance of such
quantitative studies in assessing systematic differences between local
and high-$z$ SN~Ia, but also makes a strong case for the need for higher
quality SN~Ia spectral data at high redshifts. From the first high-$z$
spectrum of an SN~Ia ever obtained (at $z=0.31$;
\citealt{Norgaard-Nielsen/etal:1989}), previous and ongoing high-$z$ 
SN~Ia surveys have gathered sufficient data for detailed quantitative
comparisons to be made between the two samples. To make the assertion
of no evolution in the SN~Ia sample with redshift, one would need a
few high-quality SN~Ia spectra, preferably with a $\sim5$ d sampling
in rest-frame phase (i.e. a $\sim1$ week sampling at $z \approx
0.5$). This could ideally be done with the {\it Hubble Space
Telescope}, but could
also be attempted with ground-based 8--10-m-class telescopes
\citep{Matheson/etal:2005}. Since the redshift uncertainty is 
the dominant source of error for our high-$z$ measurements (when
the redshift is determined from the SN itself), it is
important for future studies of absorption and emission-peak velocities in
SN~Ia to obtain a spectrum of the host galaxy along with that of the
supernova. 

The ESSENCE high-$z$ SN~Ia spectra analyzed in this paper, and
initially presented in \citet{Matheson/etal:2005}, are now publicly
available.

%%%%%%%%%%%%%%%%%%%%%%%%%%%%%%%%%%%%%%%%%%%%%%%%%%%%%%%%%%%%%%%%%%%
%%
%%   ACKNOWLEDGEMENTS
%%
%%%%%%%%%%%%%%%%%%%%%%%%%%%%%%%%%%%%%%%%%%%%%%%%%%%%%%%%%%%%%%%%%%%

\acknowledgments
The ESSENCE project is supported primarily by grants AST-0206329 and 
AST-0443378 from the U.S. National Science Foundation.
S.~B. would like to thank Nando Patat and Paolo Mazzali for useful
discussions on SN~Ia spectral features and spectral synthesis. Darrin
Casebeer's help in installing and running SYNOW is also greatly
appreciated. S.~B. acknowledges support from the International
Max-Planck Research School (IMPRS) on Astrophysics {\it via} a graduate 
fellowship.
A.V.F. is grateful for the support of NSF grant AST-0307894, and for
a Miller Research Professorship at UC Berkeley during which part of
this work was completed.\\

%%%%%%%%%%%%%%%%%%%%%%%%%%%%%%%%%%%%%%%%%%%%%%%%%%%%%%%%%%%%%%%%%%%
%%
%%   BIBLIOGRAPHY
%%
%%%%%%%%%%%%%%%%%%%%%%%%%%%%%%%%%%%%%%%%%%%%%%%%%%%%%%%%%%%%%%%%%%%

%%%%%%%%%%%%%%%%%%%%%%%%%%%%%%%%%%%%%%%%%%%%%%%%%%%%%%%%%%%%%%%%%%%
%%
%%   TABLES
%%
%%%%%%%%%%%%%%%%%%%%%%%%%%%%%%%%%%%%%%%%%%%%%%%%%%%%%%%%%%%%%%%%%%%

%%%%%%%%%%%%%%%%%%%%%%%%%%%%%%%%%%%%%%%%%%%%%%%%%%%%%%%%%%%%%%%%%%%
%
%    Table -- Local SN~Ia data
%
%%%%%%%%%%%%%%%%%%%%%%%%%%%%%%%%%%%%%%%%%%%%%%%%%%%%%%%%%%%%%%%%%%%
%
% Table of local SN~Ia data used - with references
%
\clearpage
\LongTables
\begin{landscape}
\begin{deluxetable}{llllllllc}
\tablenum{3}
\tabletypesize{\scriptsize}
\setlength{\tabcolsep}{0.03in}
\tablewidth{0pt}
%\rotate
\tablecaption{Local SN~Ia data \label{Table:localsn1adata}}
\tablehead{
\colhead{SN IAU Name} & 
\colhead{Host galaxy} & 
\colhead{$cz$} & 
\colhead{$E(B-V)$} & 
\colhead{$M_B$} &
\colhead{${\Delta m_{15}}$} & 
\colhead{HJD${_{\rm max}}$} & 
\colhead{Phases} & 
\colhead{Ref.} \\ 	
\colhead{(1)} & 
\colhead{(2)} & 
\colhead{(3)} & 
\colhead{(4)} & 
\colhead{(5)} & 
\colhead{(6)} & 
\colhead{(7)} & 
\colhead{(8)} &
\colhead{(9)}}
\startdata
1981B    & NGC~4536       & 1808                     & 0.11 (0.03)                        & $-$19.616 (0.038) & 1.10 (0.07)                        & 44670.95 (0.73) &+0,+6,+17,+20,+24                                  & (1)        \\
1986G    & NGC~5128       & 547                      & 0.50 (0.05)                        & $-$18.273 (0.113) & 1.73 (0.07)                        & 46561.36 (0.24) & $-$[5-1],+[0-3],+5                                & (2-3)      \\
1989B    & NGC~3627 (M66) & 727                      & 0.34 (0.04)                        & $-$19.537 (0.064) & 1.31 (0.07)                        & 47564.32 (0.59) & $-$6,+0,+4,+6,+[8-9],+10,+[12-15],+18             & (4-7)      \\
1990N    & NGC~4639       & 1010                     & 0.09 (0.03)                        & $-$19.704 (0.022) & 1.07 (0.05)                        & 48081.89 (0.17) & $-$[14-13],$-$[8-6],+5,+8,+[17-18]                & (8)        \\
1990O    & MCG +03-44-3   & 9193                     & 0.02 (0.03)                        & $-$19.657 (0.037) & 0.96 (0.10)                        & 48076.13 (1.02) & $-$[7-5],+0,+[20-21]	                         & (9)        \\
1991T    & NGC~4527       & 1736                     & 0.14 (0.05)                        & $-$19.694 (0.015) & 0.94 (0.05)                        & 48374.03 (0.14) & +[8-11],+[14-17],+26                              & (10-12)    \\
1991bg   & NGC~4374 (M84) & 1060                     & 0.03 (0.05)                        & $-$17.604 (0.098) & 1.93 (0.10)                        & 48603.12 (0.31) & +[1-3],+[15-16],+19,+26                           & (13-15)    \\ 
1992A    & NGC~1380       & 1877                     & 0.00 (0.02)                        & $-$19.148 (0.062) & 1.47 (0.05)                        & 48640.63 (0.19) & $-$5,+0,+3,+[5-7],+9,+12,+16,+17                  & (16)       \\
\enddata
\end{deluxetable}
\clearpage
\end{landscape}

%%%%%%%%%%%%%%%%%%%%%%%%%%%%%%%%%%%%%%%%%%%%%%%%%%%%%%%%%%%%%%%%%%%
%
%    Table -- High-z SN~Ia data
%
%%%%%%%%%%%%%%%%%%%%%%%%%%%%%%%%%%%%%%%%%%%%%%%%%%%%%%%%%%%%%%%%%%%
%
% Table of high-z SN~Ia data used 
%
\begin{deluxetable}{l l l l l l}
\tabletypesize{\scriptsize}
\tablenum{4}
\tablewidth{0pt}
\tablecaption{High-$z$ SN~Ia data \label{Table:hzsn1adata}}
\tablehead{
\colhead{IAU name} & 
\colhead{$z${\tablenotemark{a}}} & 
\colhead{$E(B-V)${\tablenotemark{b}}} & 
\colhead{\dmm{\tablenotemark{b}}} & 
\colhead{MJD${_{\rm max}}${\tablenotemark{b}}} & 
\colhead{Phases{\tablenotemark{c}}}}
\startdata
2002iy & 0.587 & 0.01 (0.03) & 0.93 (0.10) & 52592.75 (1.34) & +14            \\
2002iz & 0.428 & 0.14 (0.08) & 1.17 (0.05) & 52587.80 (1.56) & $-$1, +19      \\
2002ja & 0.33  & 0.19 (0.08) & 0.94 (0.09) & 52585.42 (1.97) & +2             \\
2002jb & 0.25  & 0.12 (0.10) & 1.10 (0.03) & 52596.01 (0.67) & $-$7           \\
2002jc & 0.52  & 0.00 (0.02) & 1.47 (0.04) & 52596.27 (0.52) & $-$6, +11      \\
2002jd & 0.32  & 0.00 (0.05) & 0.84 (0.04) & 52593.29 (0.56) & $-$3, +16, +18 \\
2002jq & 0.49  & 0.06 (0.11) & 1.48 (0.07) & 52603.47 (0.02) & +6             \\
2002js & 0.54  & 0.00 (0.03) & 1.03 (0.07) & 52600.21 (1.19) & +7             \\
\enddata					
\tablenotetext{a}{For redshifts determined from narrow lines in the
host galaxy (quoted with three decimal places), we assume an error of
0.001; for those determined {\it via} cross-correlations with local
SN~Ia spectra templates \citep{Matheson/etal:2005}, we assume an error
of 0.01.} 
\tablenotetext{b}{Light-curve parameters (and $1\sigma$ errors) output
by the fitting routine of \citet{Prieto/Rest/Suntzeff:2005}.
$E(B-V)$ values correspond to host-galaxy extinction only.}
\tablenotetext{c}{Rest-frame phase in days from rest-frame $B$-band
maximum, rounded to closest whole day.}						
\end{deluxetable}

%%%%%%%%%%%%%%%%%%%%%%%%%%%%%%%%%%%%%%%%%%%%%%%%%%%%%%%%%%%%%%%%%%%
%
%    Include Tables of vabs and vpeak measurements
%
%%%%%%%%%%%%%%%%%%%%%%%%%%%%%%%%%%%%%%%%%%%%%%%%%%%%%%%%%%%%%%%%%%%

%%%%%%%%%%%%%%%%%%%%%%%%%%%%%%%%%%%%%%%%%%%%%%%%%%%%%%%%%%%%%%%%%%%
%
%    Table -- absorption velocities in local SN~Ia
%
%%%%%%%%%%%%%%%%%%%%%%%%%%%%%%%%%%%%%%%%%%%%%%%%%%%%%%%%%%%%%%%%%%%
%
% Table of absorption velocities in local normal SN~Ia 
%
\begin{deluxetable}{l c c c c c}
\tabletypesize{\scriptsize}
\tablenum{6}
\tablewidth{0pt}
\tablecaption{Absorption velocities in local SN~Ia (10$^3$\kms)\label{Table:vabslocal}} 
\tablehead{
\colhead{Phase{\tablenotemark{a}}}      &      
\multicolumn{2}{c}{\catwo\ $\lambda3945${\tablenotemark{b}}}  & 
\colhead{\stwo\ $\lambda5454$}             & 
\colhead{\stwo\ $\lambda5640$}             & 
\colhead{\sitwo\ $\lambda6355$}            \\  
\colhead{}              &
\colhead{Blue absorption}   &
\colhead{Red/Single absorption}  &
\colhead{}              &
\colhead{}              &
\colhead{}}   
\startdata
\multicolumn{6}{c}{1981B} \\
\hline
   +0 &                         $\cdots$ &                  $-18.8 \pm 0.3$ &                  $-12.4 \pm 0.2$ &                  $-13.4 \pm 0.2$ &                  $-14.8 \pm 0.2$ \\
   +6 &                 $\cdots^\dagger$ &                 $\cdots^\dagger$ &                 $\cdots^\dagger$ &                  $-10.4 \pm 0.2$ &                  $-12.7 \pm 0.2$ \\
  +17 &                         $\cdots$ &                  $-14.3 \pm 0.4$ &                         $\cdots$ &                         $\cdots$ &                  $-11.5 \pm 0.2$ \\
  +20 &                         $\cdots$ &                  $-14.0 \pm 0.3$ &                         $\cdots$ &                         $\cdots$ &                  $-11.0 \pm 0.2$ \\
  +24 &                         $\cdots$ &                  $-14.0 \pm 0.3$ &                         $\cdots$ &                         $\cdots$ &                  $-10.5 \pm 0.2$ \\
\hline
\multicolumn{6}{c}{1986G} \\
\hline
 $-$5 &                 $\cdots^\dagger$ &                 $\cdots^\dagger$ &                  $-10.1 \pm 0.2$ &                  $-10.3 \pm 0.2$ &                  $-12.8 \pm 0.2$ \\
\enddata
\tablenotetext{$\dagger$}{Insufficient wavelength coverage.}
\tablenotetext{a}{Rest-frame days from $B$-band maximum, rounded to
closest whole day.} 
\tablenotetext{b}{\vabs\ measurements for \catwo\ \l3945 are separated
into ``blue'' and ``red'' components (when a double absorption is
present). Both ``red'' and single-absorption velocities are reported
in the same column.} 
\tablenotetext{c}{For SN~1994D there is evidence for several ``blue'' \catwo\
components \citep[see][their Fig.~3]{Hatano/etal:1999}. In the ``blue''
column we report the absorption velocity corresponding to the bluest
component.}
\end{deluxetable}

%%%%%%%%%%%%%%%%%%%%%%%%%%%%%%%%%%%%%%%%%%%%%%%%%%%%%%%%%%%%%%%%%%%
%
%    Table -- Absorption velocities in High-z SN~Ia
%
%%%%%%%%%%%%%%%%%%%%%%%%%%%%%%%%%%%%%%%%%%%%%%%%%%%%%%%%%%%%%%%%%%%
%
% Table of absorption velocities in high-z SN~Ia 
%
\begin{deluxetable}{l l c c c c c}
\tabletypesize{\scriptsize}
\tablenum{7}
\tablewidth{0pt}
\tablecaption{Absorption velocities in ESSENCE high-$z$ SN~Ia (10$^3$\kms)\label{Table:vabshighz}} 
\tablehead{
\colhead{IAU name}                             &
\colhead{Phase{\tablenotemark{a}}}             &      
\multicolumn{2}{c}{\catwo\ $\lambda3945${\tablenotemark{b}}}  & 
\colhead{\stwo\ $\lambda5454$}             & 
\colhead{\stwo\ $\lambda5640$}             & 
\colhead{\sitwo\ $\lambda6355$}            \\  
\colhead{}              &
\colhead{}              &
\colhead{Blue absorption}            &
\colhead{Red/Single absorption}  &
\colhead{}              &
\colhead{}              &
\colhead{}}   
\startdata
2002iy$^\ddagger$ &   +14 &                         $\cdots$ &                  $-13.0 \pm 1.5$ &                         $\cdots$ &                         $\cdots$ &                         $\cdots$ \\
           2002iz &  $-$2 &                         $\cdots$ &                  $-23.7 \pm 1.1$ &                         $\cdots$ &                         $\cdots$ &                  $-12.2 \pm 0.9$ \\
2002iz$^\ddagger$ &   +18 &                         $\cdots$ &                  $-13.8 \pm 1.5$ &                         $\cdots$ &                         $\cdots$ &                         $\cdots$ \\
           2002ja &    +1 &                         $\cdots$ &                         $\cdots$ &                         $\cdots$ &                         $\cdots$ &                  $-14.6 \pm 3.1$ \\
           2002jb &  $-$7 &                         $\cdots$ &                  $-17.2 \pm 3.2$ &                         $\cdots$ &                         $\cdots$ &                  $-11.2 \pm 3.0$ \\
           2002jc &  $-$6 &                         $\cdots$ &                  $-21.1 \pm 3.2$ &                         $\cdots$ &                         $\cdots$ &                         $\cdots$ \\
           2002jc &   +11 &                         $\cdots$ &                  $-11.8 \pm 3.2$ &                         $\cdots$ &                         $\cdots$ &                 $\cdots^\dagger$ \\
           2002jd &  $-$3 &                         $\cdots$ &                  $-16.9 \pm 3.0$ &                  $-10.8 \pm 3.0$ &                  $-12.6 \pm 3.0$ &                  $-12.0 \pm 3.1$ \\
\enddata
\tablenotetext{$\dagger$}{Insufficient wavelength coverage.}
\tablenotetext{$\ddagger$}{VLT spectra reduced with the 2D restoration 
method of \citet{Blondin/etal:2005}.}
\tablenotetext{a}{Rest-frame days from $B$-band maximum, rounded to
closest whole day.} 
\tablenotetext{b}{\vabs\ measurements for \catwo\ \l3945 are separated
into ``blue'' and ``red'' components (when a double absorption is present). Both
``red'' and single-absorption velocities are reported in the same column.}
\end{deluxetable}

%%%%%%%%%%%%%%%%%%%%%%%%%%%%%%%%%%%%%%%%%%%%%%%%%%%%%%%%%%%%%%%%%%%
%
%    Table -- emission-peak velocities in local SN~Ia
%
%%%%%%%%%%%%%%%%%%%%%%%%%%%%%%%%%%%%%%%%%%%%%%%%%%%%%%%%%%%%%%%%%%%
%
% Table of emission-peak velocities in local normal SN~Ia 
%
\begin{deluxetable}{l c c c c}
\tabletypesize{\scriptsize}
\tablenum{8}
\tablewidth{0pt}
\tablecaption{Emission-peak velocities in local SN~Ia (10$^3$\kms)\label{Table:vpeaklocal}} 
\tablehead{
\colhead{Phase{\tablenotemark{a}}}      &      
\colhead{\catwo\ $\lambda3945$}  & 
\colhead{\stwo\ $\lambda5454$}             & 
\colhead{\stwo\ $\lambda5640$}             & 
\colhead{\sitwo\ $\lambda6355$}}   
\startdata
\multicolumn{5}{c}{1981B} \\
\hline
   +0 &              $-5.4 \pm 0.3$ &              $-7.5 \pm 0.2$ &              $-6.7 \pm 0.2$ &              $-3.8 \pm 0.2$ \\
   +6 &            $\cdots^\dagger$ &            $\cdots^\dagger$ &              $-5.0 \pm 0.2$ &              $-1.7 \pm 0.2$ \\
  +17 &              $-1.4 \pm 0.4$ &                    $\cdots$ &                    $\cdots$ &                    $\cdots$ \\
  +20 &              $-1.1 \pm 0.3$ &                    $\cdots$ &                    $\cdots$ &                    $\cdots$ \\
  +24 &              $-0.2 \pm 0.3$ &                    $\cdots$ &                    $\cdots$ &                    $\cdots$ \\
\hline
\multicolumn{5}{c}{1986G} \\
\hline
 $-$5 &            $\cdots^\dagger$ &              $-5.4 \pm 0.2$ &              $-1.7 \pm 0.2$ &                    $\cdots$ \\
\enddata
\tablenotetext{$\dagger$}{Insufficient wavelength coverage.}
\tablenotetext{a}{Rest-frame days from $B$-band maximum, rounded to
closest whole day.}
\end{deluxetable}

%%%%%%%%%%%%%%%%%%%%%%%%%%%%%%%%%%%%%%%%%%%%%%%%%%%%%%%%%%%%%%%%%%%
%
%    Table -- Emission-Peak velocities in High-z SN~Ia
%
%%%%%%%%%%%%%%%%%%%%%%%%%%%%%%%%%%%%%%%%%%%%%%%%%%%%%%%%%%%%%%%%%%%
%
% Table of emission-peak velocities in high-z SN~Ia 
%
\begin{deluxetable}{l c c c c c}
\tabletypesize{\scriptsize}
\tablenum{9}
\tablewidth{0pt}
\tablecaption{Emission-peak velocities in ESSENCE high-$z$ SN~Ia (10$^3$\kms)\label{Table:vpeakhighz}} 
\tablehead{
\colhead{IAU name}                             &
\colhead{Phase{\tablenotemark{a}}}             &      
\colhead{\catwo\ $\lambda3945$}  & 
\colhead{\stwo\ $\lambda5454$}             & 
\colhead{\stwo\ $\lambda5640$}             & 
\colhead{\sitwo\ $\lambda6355$}}   
\startdata
2002iy$^\ddagger$ &   +14 &              $-3.3 \pm 1.5$ &                    $\cdots$ &                    $\cdots$ &                    $\cdots$ \\
           2002iz &  $-$2 &              $-3.5 \pm 1.1$ &                    $\cdots$ &                    $\cdots$ &              $-5.7 \pm 0.9$ \\
2002iz$^\ddagger$ &   +18 &              $-3.8 \pm 1.5$ &                    $\cdots$ &                    $\cdots$ &                    $\cdots$ \\
           2002ja &    +1 &                    $\cdots$ &                    $\cdots$ &                    $\cdots$ &              $-4.8 \pm 3.1$ \\
           2002jb &  $-$7 &                    $\cdots$ &                    $\cdots$ &                    $\cdots$ &              $-1.5 \pm 3.0$ \\
           2002jc &  $-$6 &              $-5.2 \pm 3.2$ &                    $\cdots$ &                    $\cdots$ &                    $\cdots$ \\
           2002jd &  $-$3 &              $-3.6 \pm 3.0$ &              $-6.4 \pm 3.0$ &              $-6.4 \pm 3.0$ &              $-4.2 \pm 3.1$ \\
2002jd$^\ddagger$ &   +16 &              $-1.9 \pm 3.3$ &                    $\cdots$ &                    $\cdots$ &              $-4.0 \pm 3.3$ \\
\enddata
\tablenotetext{$\dagger$}{Insufficient wavelength coverage.}
\tablenotetext{$\ddagger$}{VLT spectra reduced with the 2D restoration 
method of \citet{Blondin/etal:2005}.}
\tablenotetext{a}{Rest-frame days from $B$-band maximum, rounded to
closest whole day.} 
\end{deluxetable}

%%%%%%%%%%%%%%%%%%%%%%%%%%%%%%%%%%%%%%%%%%%%%%%%%%%%%%%%%%%%%%%%%%%
%%
%%   THE END
%%
%%%%%%%%%%%%%%%%%%%%%%%%%%%%%%%%%%%%%%%%%%%%%%%%%%%%%%%%%%%%%%%%%%%

\end{document}